\documentclass[aps,prl,twocolumn,superscriptaddress,citeautoscript,amssymb,reprint,showpacs,floatfix]{revtex4-2}
\usepackage[ansinew]{inputenc}
\usepackage[T1]{fontenc}
\usepackage{float}
\usepackage{amsmath}
\usepackage{mathptmx}
\usepackage{mathrsfs}
\usepackage[english]{babel}
\usepackage{graphicx}
\usepackage{hyperref}
\usepackage[usenames,dvipsnames]{xcolor}
\usepackage{natbib}
\usepackage{makecell}
\usepackage[per-mode=reciprocal, range-units = single, range-phrase=\ -\ , exponent-product = \cdot , separate-uncertainty = true, multi-part-units=single , quotient-mode = fraction]{siunitx}

\usepackage{blindtext}
\usepackage{physics}
\usepackage[normalem]{ulem}

\newcommand{\ybalo}{YbAlO$_3$}

\newcommand{\Bacovo}{BaCo$_2$V$_2$O$_8$}
\newcommand{\Srcovo}{SrCo$_2$V$_2$O$_8$}
\newcommand{\be}{\begin{equation}}
\newcommand{\ee}{\end{equation}}
\newcommand{\bea}{\begin{eqnarray}}
\newcommand{\eea}{\end{eqnarray}}

\def\SP#1{\textsuperscript{{#1}}}

\begin{document}

\author{P. Mokhtari}
\thanks{Present address: Department of Applied Physics and Quantum-Phase Electronics Center, The University of Tokyo, Bunkyo-ku, Tokyo 113-8656, Japan}
\affiliation{Department of Physics, Technical University of Munich, 85748 Garching, Germany}
\affiliation{Max Planck Institute for Chemical Physics of Solids, 01187 Dresden, Germany}
\affiliation{Technische Universit\"at Dresden, 01062 Dresden, Germany}

\author{S. Galeski}
\affiliation{Max Planck Institute for Chemical Physics of Solids, 01187 Dresden, Germany}
\affiliation{Physikalisches Institut, Universitat Bonn, Nussallee 12, 53115 Bonn, Germany}

\author{U. Stockert}
\affiliation{Technische Universit\"at Dresden, 01062 Dresden, Germany}

\author{S.~E.~Nikitin}
\affiliation{Laboratory for Neutron Scattering and Imaging, PSI Center for Neutron and Muon Sciences, Paul Scherrer Institut, CH-5232 Villigen-PSI, Switzerland}

\author{R. Wawrzy\'nczak}
\affiliation{Max Planck Institute for Chemical Physics of Solids, 01187 Dresden, Germany}

\author{R. K{\"u}chler}
\affiliation{Max Planck Institute for Chemical Physics of Solids, 01187 Dresden, Germany}

\author{M. Brando}
\affiliation{Max Planck Institute for Chemical Physics of Solids, 01187 Dresden, Germany}

\author{L. Vasylechko}
\affiliation{Lviv Polytechnic National University, Lviv 79013, Ukraine}

\author{O. A. Starykh}
\thanks{correspondence should be addressed to oleg.starykh@utah.edu  and elena.hassinger@tu-dresden.de}
\affiliation{Department of Physics and Astronomy, University of Utah, USA}

\author{E. Hassinger}
\thanks{correspondence should be addressed to oleg.starykh@utah.edu  and elena.hassinger@tu-dresden.de}
\affiliation{Technische Universit\"at Dresden, 01062 Dresden, Germany}
\affiliation{Max Planck Institute for Chemical Physics of Solids, 01187 Dresden, Germany}

\title{
1/5 and 1/3 magnetization plateaux in the spin 1/2 chain system YbAlO$_3$}
\date{\today}

\begin{abstract}

Quasi-one-dimensional magnets can host an ordered longitudinal spin-density wave state (LSDW) in magnetic field at low temperature, when longitudinal correlations are strengthened by Ising anisotropies. In the S = 1/2 Heisenberg antiferromagnet YbAlO$_3$ this happens via Ising-like interchain interactions. Here, we report the first experimental observation of magnetization plateaux at 1/5 and 1/3 of the saturation value via thermal transport and magnetostriction measurements in YbAlO$_3$.
We present a phenomenological theory of the plateau states that describes them as islands of commensurability within an otherwise incommensurate LSDW phase and explains their relative positions within the LSDW phase and their relative extent in a magnetic field. Notably, the plateaux are stabilised by ferromagnetic interchain interactions in \ybalo~and consistently are absent in other quasi-1D magnets such as \Bacovo~with antiferromagnetic interchain interactions. We also report a small, step-like increase of the magnetostriction coefficient, indicating a weak phase transition of unknown origin in the high-field phase just below the saturation.
\end{abstract}

\maketitle

\textit{Introduction}---
Low-dimensional quantum magnets support a large variety of exotic quantum states, such as quantum spin liquids~\cite{Savary2017}, magnetization plateaux, or nematic states that are induced by quantum fluctuations \cite{Starykh2015}. Quasi-one-dimensional magnets are vital in this field since they are generally well understood theoretically~\cite{Giamarchi_book}. In recent years, this enhanced theoretical understanding has been translated into a number of spectacular experimental observations that include a realization of the quantum integrable model with extended E$_8$ symmetry~\cite{zamolodchikov1989,coldea2010,Wu2014}, many-body string excitations~\cite{Wang2018,Wang2019,Yang2023}, and repulsively bound magnon states~\cite{Halati2023,Wang2024}. Most of these observations are based on spin-1/2 chain materials with pronounced Ising anisotropies, such as CoNb$_2$O$_6$~\cite{coldea2010}, \Srcovo~ \cite{Wang2018}, and \Bacovo~ \cite{Wang2019}. These materials are more complex than minimalistic theoretical models inspired by them, and important details of their magnetic field $B$ - temperature $T$ phase diagrams remain to be understood~\cite{Klanjsek2015,Takayoshi2023}.

Here, we report the experimental discovery of multiple magnetization plateaux in another quasi-one-dimensional magnet with the Ising motif, YbAlO$_3$. In contrast with the examples listed above, in YbAlO$_3$ the exchange interaction between spins within the chain is of Heisenberg kind, while that between the spins from neighboring chains is dominantly Ising-like~\cite{wu_antiferromagnetic_2019,Wu-Zaliznyak2016}. The latter feature originates from the dipole-dipole nature of the interchain interactions~\cite{wu_tomonagaluttinger_2019,wu_antiferromagnetic_2019}. It provides a novel route to the incommensurate longitudinal spin-density wave (LSDW) phase, a state that, in many respects, is similar to an itinerant charge-neutral conductor with the magnetic field-dependent Fermi-momenta $k_{\rm F} = \pi (1 \pm 2M)/2 = \pi (1 \pm m)/2$, where $m=M/M_s$ is the magnetization per site $M = \langle S^z\rangle$ normalized by the saturation value $M_{\rm s}=1/2$. 
We find that LSDW hosts two magnetization plateaux at 1/5 and 1/3 of the saturation value. 
While the 1/3 plateau has been previously observed in neutron-scattering and magnetization studies ~\cite{wu_tomonagaluttinger_2019,nikitin_multiple_2021}, the plateau at 1/5 is the new result. Notably, both plateaux reported here are detected via the thermal transport and magnetostriction measurements. The measurements are done at sub-Kelvin temperatures $\sim 0.1$K. We find that the intrinsic spin thermal conductivity is larger than the phonon contribution and can be well separated from it. We present field-theoretic and symmetry-based arguments in favor of the magnetization-plateau stabilization by the ferromagnetic interchain interactions. Furthermore, we detect a phase transition of yet unknown origin not far below the quantum phase transition to the field-polarised state. Our manuscript provides new information on the magnetic phase diagram of YbAlO$_3$ and motivates further numerical studies of its microscopic spin Hamiltonian.

\begin{figure}
    \center{\includegraphics[width=\columnwidth]{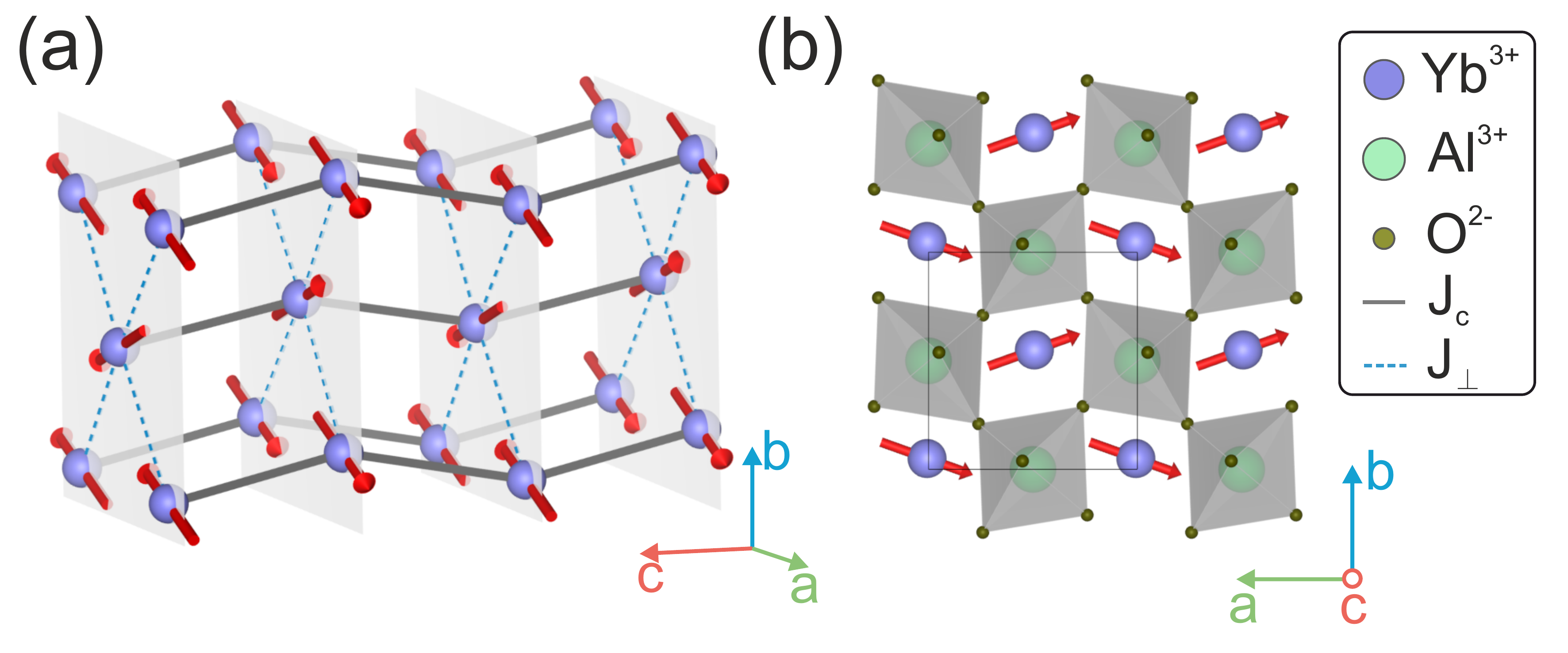}}
     \caption{
     (a)~Sketch of the magnetic structure showing only Yb ions and two relevant exchange interactions, $J_c$ and $J_{\perp}$. 
     (b)~Crystal structure of \ybalo\ viewed from [001] direction.}
     \label{fig:mag_Hamiltonian}
\end{figure}

\textit{Formation of LSDW in YbAlO$_3$}---
YbAlO$_3$ is a  rare-earth-based insulator with an orthorhombically distorted perovskite structure as represented in Fig.~\ref{fig:mag_Hamiltonian}b with room-temperature lattice constants $a=5.126$\,\AA, $b=5.331$\,\AA, and $c=7.313$\,\AA~(in conventional $Pbnm$ notation)~\cite{Buryy_growth_2010}. Due to the crystal electric fields, the Yb $J = 7/2$ multiplet splits into four doublets, with the lowest-energy doublet well separated from the higher CEF levels, leading to an effective $S=1/2$ system~\cite{wu_antiferromagnetic_2019}. The Yb moments have a strong uniaxial g-factor anisotropy with a local easy axis oriented within the $ab$ plane with an angle of $\pm 23.5^\circ$  away from the $a$ axis [Fig.~\ref{fig:mag_Hamiltonian}a]~\cite{wu_tomonagaluttinger_2019,ehlers_esp_2022} and the $g$ factors $g_\parallel = 7.6$ (so that the full magnetic moment is $g_\parallel \mu_B/2=3.8\,\mu_\mathrm{B}/\mathrm{Yb}$\cite{RADHAKRISHNA1981}), much larger than $g_\perp~\approx~0.46$\cite{wu_tomonagaluttinger_2019,wu_antiferromagnetic_2019,RADHAKRISHNA1981,ehlers_esp_2022}. 
Thus, the crystallographic $a$ axis is the direction with the highest and equal $g$ factor for both Yb sites in the crystal structure. In our study, the magnetic field $B$ is applied along this $a$ axis.

The spin chains run along the $c$ axis and are well described by the isotropic Heisenberg intrachain exchange coupling $J_\textrm{c} = 2.4$\,K \cite{wu_tomonagaluttinger_2019,wu_antiferromagnetic_2019,nikitin_dimers_2020}. 
Recent neutron scattering reveals a gapless spinon continuum at 1\,K, and an AF state appears at 0.88\,K due to interchain interactions with effective $J_\perp \approx -0.2J_\mathrm{c}= - 0.5$\,K~\cite{RADHAKRISHNA1981,wu_tomonagaluttinger_2019,fan_role_2020}. The interchain interactions are likely ferromagnetic and of dipolar origin, leading to an A-type order with distorted ferromagnetic arrangement in the planes perpendicular to the $c$ axis. The magnetic structure and dominant exchange coupling parameters are shown in Fig.~\ref{fig:mag_Hamiltonian}a. Surprisingly, the phase diagram of YbAlO$_3$ in magnetic field resembles that of the Ising materials \Bacovo~and \Srcovo~in which, after suppression of the AF order at $B_\textrm{c}$, an incommensurate LSDW order is established because the longitudinal spin-spin correlations are strengthened by the Ising character of the exchange interactions~\cite{Okunishi2007,Takayoshi2023}. 
Theoretical studies inspired by YbAlO$_3$ suggest that even for isotropic Heisenberg chains, Ising-like interchain interactions can also stabilize the LSDW \cite{fan_role_2020,wu_tomonagaluttinger_2019,fan_quantumcricality_2020, Agrapidis_incommensurate_2019}.

\begin{figure}
\centering
     \includegraphics[width=\columnwidth]{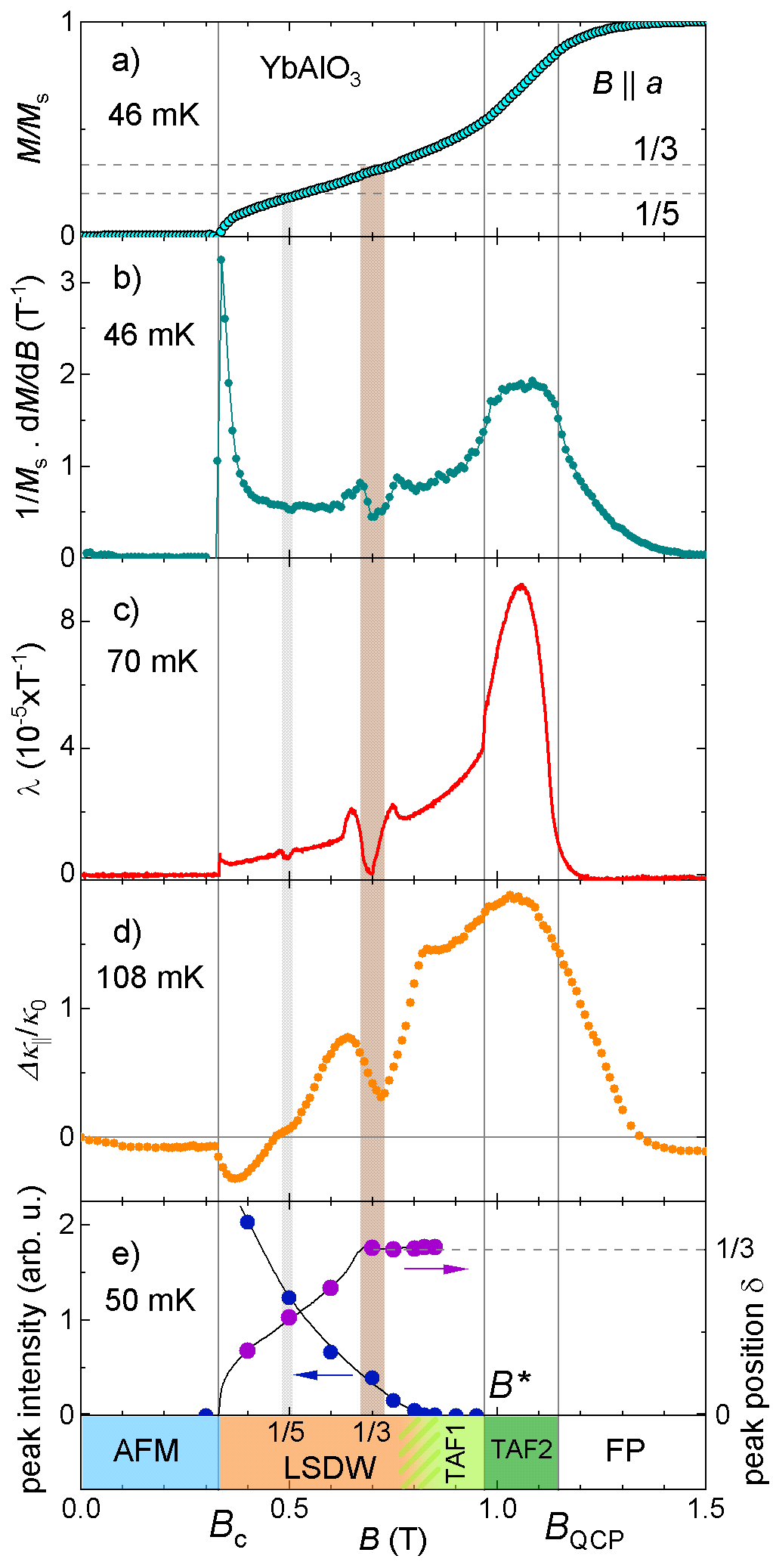}
     \caption{Field dependence of different quantities in YbAlO$_3$ at low temperature for $H\parallel a$. Sharp anomalies occur at $B_\mathrm{c}$, $B^\ast$ and at the fields where the magnetization reaches 1/5 $M_\mathrm{s}$ as well as 1/3 $M_\mathrm{s}$ in all quantities: normalised magnetization $M/M_\mathrm{s}$ (a), and its derivative (b), magnetostriction coefficient $\lambda$ (c), and thermal conductivity $\kappa$, here shown as the conductivity change normalised by the zero-field value (d). The 1/3 plateau is also evidenced by a constant position $\mathbf{Q} = (0,0,Q)$ with $Q = \pi(1+\delta)$ and $\delta = 1/3$ of the magnetic Bragg peak in neutron scattering (e, right axis) associated with the LSDW. The Bragg peak intensity is finite and the LSDW state persists up to 0.85\,T but becomes very small for fields above 0.75\,T (e, left axis).}
     \label{fig:plateaux}
\end{figure}

Prime evidence for the LSDW in YbAlO$_3$ comes from the comparison of the Bragg peak position with the magnetization showing exactly the expected behavior~\cite{wu_tomonagaluttinger_2019,nikitin_multiple_2021}: 
The propagation wavevector $\mathbf{Q} = (0,0,Q)$, where $Q=\pi(1\pm\delta)$ and the incommensurability $\delta = m = 2M$ scales with the magnetization and is aligned along the chain direction corresponding to the $c$ axis of the crystal, whereas the magnetic moments point along the easy axis direction. Note that $Q = \pm 2k_F$ up to the lattice momentum $2\pi$, characteristic of the LSDW state.
As observed previously and reproduced here, Fig.~\ref{fig:plateaux}(a,b) shows the magnetization plateau state for $M_\textrm{s}/3$ for which $Q/\pi$ locks into the commensurate position $\delta = 1/3$ in Fig.~\ref{fig:plateaux}e \cite{wu_tomonagaluttinger_2019,nikitin_multiple_2021}. 

\textit{Experimental observation of magnetization plateaux}---
Figure ~\ref{fig:plateaux} represents the experimental results of different probes versus field up to the field-polarised state. Additional details are provided in the Sec. S1 in SM~\cite{sm1}. For each probe, the data is taken at the lowest temperature available.
Panels a and b show, respectively, the magnetization normalized to the saturation value $M/M_\mathrm{s}$ and its derivative. As expected, the magnetization rises sharply from zero at $B_{\rm c} = 0.32$\,T and reaches the saturation value $M_\mathrm{s}$ at $\approx 1.4$\,T at this temperature. The plateau at $\frac{1}{3}M_\mathrm{s}$ is clearly visible at a magnetic field of 0.7\,T. The field of the quantum critical point $B_\mathrm{QCP} = 1.15$\,T was previously derived~\cite{wu_antiferromagnetic_2019}.\\
The magnetostriction coefficient, defined as $\lambda = \frac{1}{L_0} \dv{L}{B}$, where \(L_0\) is the sample dimension, is shown in panel c. Magnetostriction is a thermodynamic bulk probe that is sensitive to magnetoelastic coupling \cite{Zapf2008,Miyata2021}. The signal here is similar to the magnetic susceptibility as seen by comparing panels c and b ($\Delta L/L_0$ is shown in Fig.~\ref{fig:magnetostriction_raw} in \cite{sm1}). The $1/3$ plateau is visible in $\lambda$ as a v-shaped anomaly analogous to the signature in the derivative of the magnetization. Being a very sensitive technique, it also resolves a second smaller v-shaped anomaly at 0.5\,T. As indicated by a dashed horizontal line in panel a, this corresponds to the field where the magnetization reaches $\frac{1}{5}M_\mathrm{s}$. Knowing the presence of this anomaly, one can also identify a corresponding small v-shaped signature in the derivative of the magnetization, panel b. To the best of our knowledge, this is the first time a 1/5 plateau has been observed in a quantum magnet.

Another anomaly in the magnetostriction is detected at $B^\ast=0.96$\,T. At this field, the magnetostriction coefficient shows a small but clear 
jump to a higher value, an unambiguous  
indication of a weak second-order phase transition, where the slope of the magnetization (panel b) also strongly increases to roughly double the value, but less sharply.

We now turn to the thermal conductivity $\kappa$, which also contains signatures of all the transitions described above, even though all of the anomalies are broader in the field. Plotted in Fig.~\ref{fig:plateaux}d is the thermal conductivity along the chain direction as a function of magnetic field $B$ relative to its value at zero magnetic field, $\Delta \kappa/{\kappa_0} = [\kappa(B) - \kappa_{0}]/\kappa_{0}$, where $\kappa_{0}= \kappa (B = \SI{0}{})$ at constant low temperature $T=108$\,mK. 
    
In general, thermal conductivity can give important information on the heat carrying excitations in low-dimensional quantum magnets \cite{Hess2019}. For a magnetic insulator such as \ybalo, phonons and magnetic excitations both contribute to the heat transport, so that $\kappa=\kappa_{\rm ph}+\kappa_{\rm mag}$. For each heat carrier, different scattering mechanisms contribute to the scattering rate. Interactions between the two kinds of heat carriers induce correlations between $\kappa_{\rm ph}$ and $\kappa_{\rm mag}$ by reducing both contributions relative to the non-interacting limit.
In our data, the low-field ($\kappa (B =0)$) and the high-field values of the thermal conductivity at the given temperature agree with each other, as Fig.\ref{fig:plateaux}d shows. This is because magnetic excitations are gapped in both limits. In the low-field limit, $B < B_{\rm c}$, the system is in the AFM Ising ordered phase, where the gap in the excitation spectrum is estimated as $T_{\rm gap}=$0.3\,meV/k$_B$ = 3.5\,K see Fig.3b in \cite{wu_tomonagaluttinger_2019}. In the high-field limit, $B > 1.5$\, T, the spin gap is controlled by the magnetic field. It thus follows that for $T \ll T_{\rm gap}$ there are no magnetic excitations in these magnetic field regions and $\kappa (B =0)$ represents the upper limit of the phonon contribution to the thermal conductivity. Conversely, the field-induced increase of $\kappa(B)$ in the intermediate LSDW-TAF field region $B_c < B < B_{\rm QCP}$ represents the {\em magnetic} contribution. Moreover, given the detrimental role of the phonon-magnon scattering, $\Delta \kappa$ introduced above represents a {\em lower} bound of the magnetic thermal conductivity $\kappa_\mathrm{mag} \geq \Delta \kappa$. (A more detailed discussion of thermal conductivity is presented in the forthcoming publications~\cite{Mokhtari2025a,Mokhtari2025}.)

This observation explains the high sensitivity of $\Delta \kappa$ in Fig.\ref{fig:plateaux}d to the magnetic field. As we argue below, the reduction of $\Delta \kappa$ inside the $1/5$ and $1/3$ plateau phases relative to the increasing $\kappa$ for the adjacent field regions has to do with the opening of the spin gap inside these {\em commensurate} SDW states. Such a gap leads to a decrease of the magnetic heat-carrier density and, as a result, a dip in  
the magnetic thermal conductivity.

\begin{figure}
     \includegraphics[width=1\columnwidth]{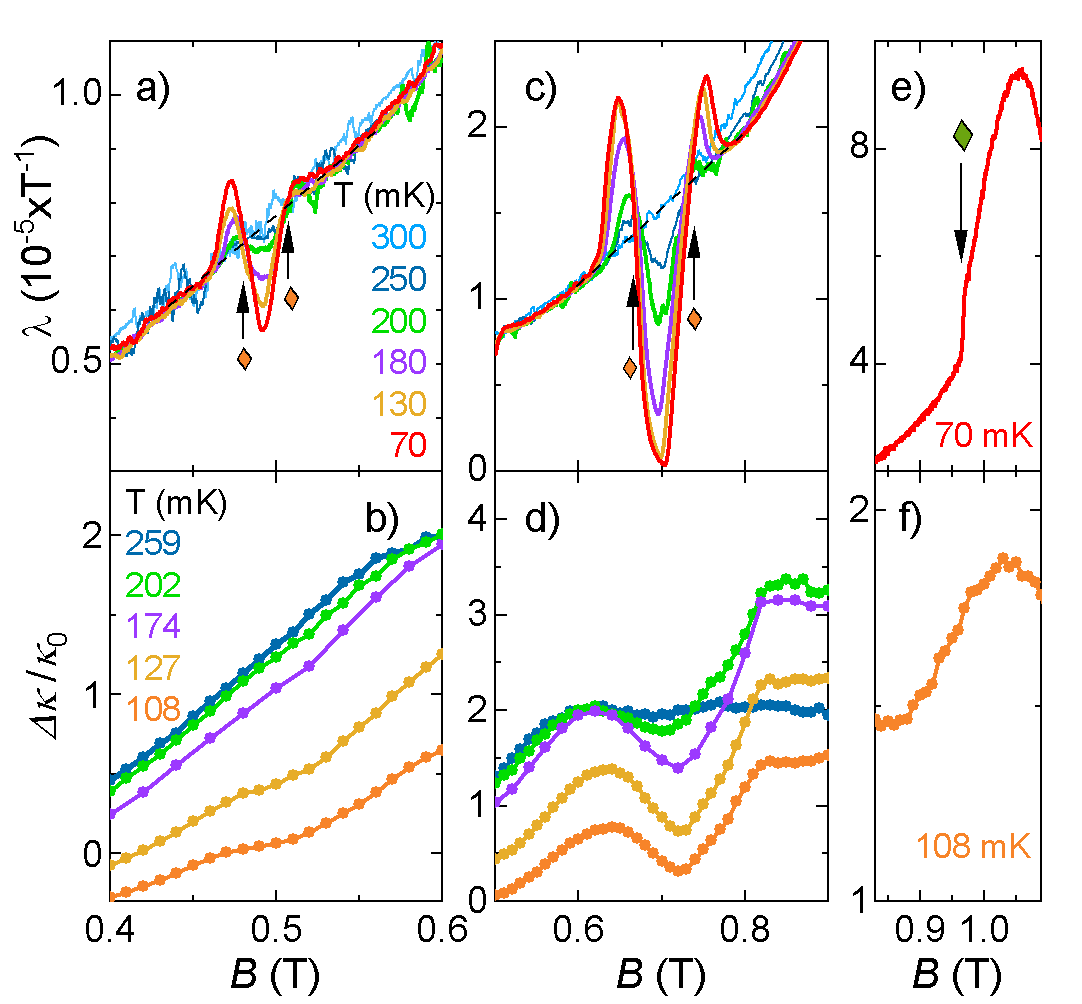}
   \caption{$T$ evolution of anomalies in magnetostriction and thermal conductivity zoomed around the 1/5 (a,b) and 1/3 (c,d) plateaux. The plateau region was defined as the field region in which the magnetostriction lies below the high-temperature curve as indicated by the arrows and orange symbols for the curve at 70\,mK. (e,f) show a zoom at the phase transition at $B^*=0.96$\,T at lowest measured temperature indicated by an arrow and a green symbol in (e).}
    \label{fig:temp_evol}
\end{figure}

Our high-quality data allow for quantitative analysis of 
the temperature dependence of the anomalies in the plateau states in Fig.~\ref{fig:temp_evol}a-d.
Based on the precise magnetostriction data, we define the width of the plateaux similar as in Ref.~\cite{RanjithPRB2019} (see Fig.~\ref{fig:temp_evol}). For both plateau states, the width remains constant with temperature. The width of the 1/5 plateau is $(26\pm 3)$\,mT and roughly 0.4 times the width of the 1/3 plateau with $(60\pm 5)$\,mT. The anomalies in the curves are visible up to $T_\mathrm{N}$ (see a more detailed evaluation of the latter in Sec. S2 in SM\cite{sm1}).
From the thermal conductivity we can estimate the size of the gap in the 1/3 plateau state as given in Sec. S3 in SM~\cite{sm1} where $\Delta_{1/3}/k_B \approx 0.19$\,K.


\begin{figure}
     \includegraphics[width=1\columnwidth]{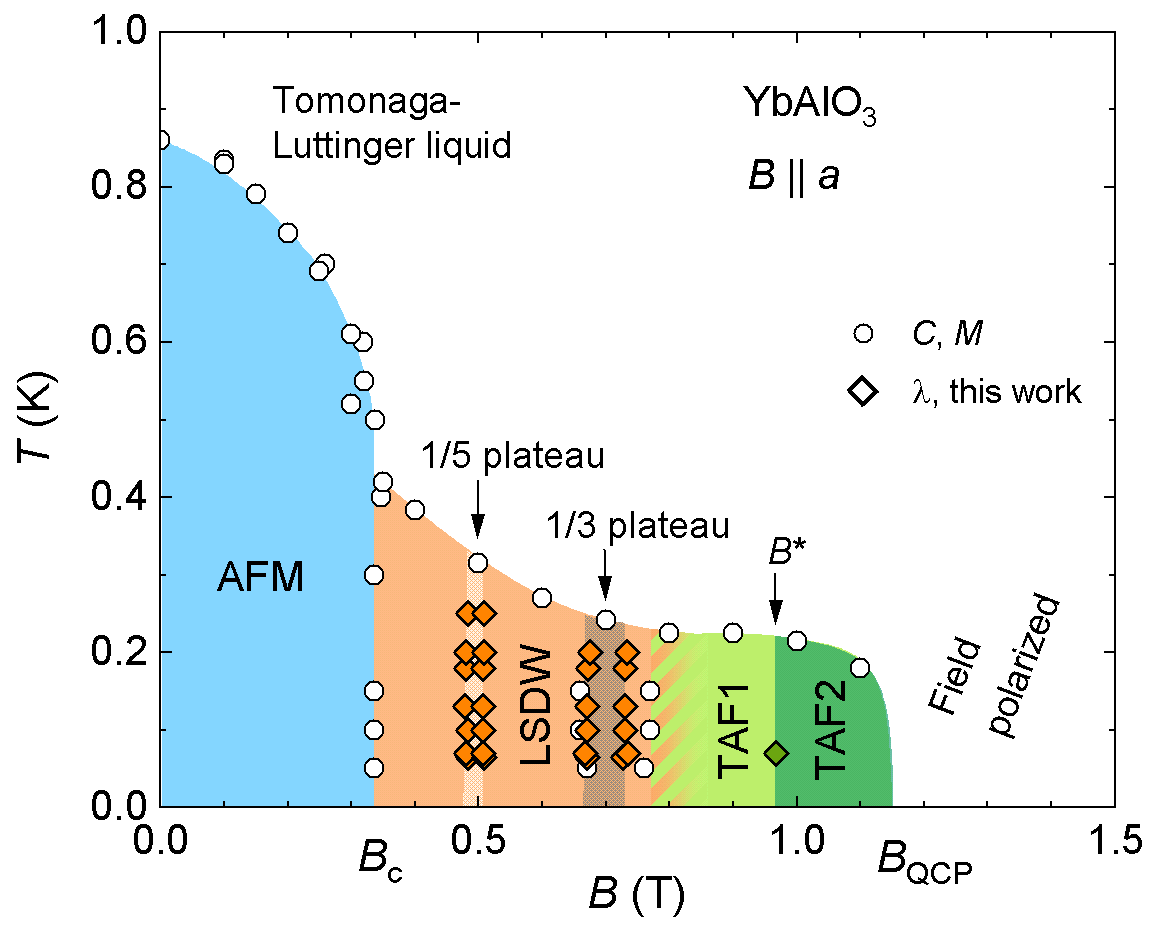}
   \caption{The phase diagram of YbAlO$_3$. White points are from specific heat $C$ and magnetization $M$ in reference \cite{wu_tomonagaluttinger_2019}. Orange and green points are from the magnetostriction $\lambda$ in this work showing the two plateau regions appearing within the LSDW state and the phase transition at $B^\ast = 0.96$\,T. See text for details.
   }
     \label{fig:phasediagram}
\end{figure}

\textit{Theoretical analysis of the plateau states}---
We now summarize key points of the theoretical analysis of plateau phases which is presented in Sec. S4 in SM~\cite{sm1}. Magnetization plateau states represent {\em commensurate} version of the LSDW phase. Once the latter is stabilized by the magnetic field and interchain spin interactions, the plateaux are bound to happen as the LSDW ordering wavevector $q = 2k_{\rm F}$ continuously scans the interval from $\pi$ at $M=0$ to $0$ at $M=1/2$. In the process, it passes through fractional values $\nu/k$ ($\nu$ and $k$ are {\em integers}) of the reciprocal lattice vector $2\pi$. For each of these occurrences, there exists a symmetry-allowed {\em umklapp} interaction that involves $k$ spins and changes the total momentum of the spin subsystem by $2\pi \nu$, i.e. by $\nu$ units of the lattice momentum. For a single Heisenberg spin chain, such multi-particle interactions are highly irrelevant and do not affect the magnetization~\cite{Giamarchi_book}. However, in the {\em ordered} three-dimensional LSDW phase, they do produce plateaux at magnetization $M_{\nu,k} = (1 - 2\nu/k)/2$ {\em provided} that the corresponding $k$-th order umklapp interaction also minimizes the much stronger interchain interaction (which is the reason for the LSDW phase in the first place). Finally, the width of the plateau (in the magnetic field) is exponentially narrow in $k^2$, which strongly favors plateaux with the smallest possible $k$ values. The detailed phenomenological analysis in Sec. S4 in~\cite{sm1} shows that ferromagnetic interchain interaction favors {\em odd}-$k$ umklapps. Together with the experimental restriction that the LSDW phase occupies a finite magnetization interval $0 < 2M \leq 1/3$ (see the section on the phase diagram below), one concludes that plateaux at $M_{1,3}$ and $M_{2,5}$ are the most prominent ones satisfying all the requirements. Moreover, the $M_{2,5}$ plateau is narrower than the $M_{1,3}$ one, precisely as the experiment shows. 
Theoretically, the next most stable plateau in the available magnetization range is $M_{3,7}$, at $1/7$ of $M_s$. The arguments above, together with the fact that $7^2/5^2 \approx 2$, predict that it should be much narrower than the already tiny $M_{2,5}$ feature, explaining its absence in our data. 

It is worth noting here that the AFM phase itself, in fact, is a {\em zero} magnetization plateau, $M_{1,2}$ in our notations. Unlike all other plateaux discussed above, it is an {\em even}-$k$ ($k=2$) state. Correspondingly, it will be present even if the interchain interaction is antiferromagnetic. Being the smallest-$k$ plateau, it is, in agreement with the theory, the widest one in the magnetic field. The AFM-LSDW transition is, therefore, of the commensurate-incommensurate (C-IC) kind. A sharp variation of the ordering wave vector $Q$ with $B$ in Fig.\ref{fig:plateaux}e, where $Q$ deviates from its commensurate $\pi$ value with an infinite slope, is a clear experimental manifestation of the C-IC physics \cite{sm1}.

\textit{The phase diagram}---
Our findings, together with the previously available data, are summarized in the phase diagram in Fig.~\ref{fig:phasediagram}. The phase diagram is guided by theoretical studies \cite{fan_role_2020,wu_tomonagaluttinger_2019,fan_quantumcricality_2020, Agrapidis_incommensurate_2019} which suggest the following sequence of the phases: Ising AFM - LSDW - TAF - FP. Here, the transverse antiferromagnetic phase (TAF) denotes a commensurate state with $Q = \pi$ and staggered spin order in the plane perpendicular to the easy axis. Neutron scattering detection of this high-field phase is severely complicated by the high anisotropy of the $g$ factor in YbAlO$_3$. Compared to the longitudinal signal, any transverse signal is reduced by a factor $(g_\parallel/g_\perp)^2 \approx 273$. FP denotes the field-polarized ferromagnetic phase.

Based on the data, LSDW phase extends from $B_c$ to about $0.75$\, T corresponding roughly to the upper end of the 1/3 plateau given that the upper end varies with measurement technique and definition. While the LSDW Bragg peak in Fig.~\ref{fig:plateaux}e persists beyond this field, its intensity drops strongly for $B>0.75$\, T. Given the different symmetries of the LSDW and TAF phases and their respective ordering wave vectors $Q$, the transition between them is likely of the 1st order. This explains the persistence of the LSDW Bragg peak into a coexistence region between 0.75\,T and 0.85\,T. The change from LSDW to TAF is also seen in the flattening of the critical temperature $T_N$ curve in Fig.~\ref{fig:phasediagram}. Fig.~\ref{figsm:001_vs_field} shows the elastic Bragg signal from the TAF phase in the interval from $0.75$\, T to the saturation. We note that this analysis is slightly complicated by the  presence of the small twin crystalline in the studied sample, as is discussed in detail in Sec. S5 in SM~\cite{sm1}. It also must be noted that the very existence of the TAF phase requires some interaction between {\em transverse} (with respect to the field) components of spins on neighboring chain \cite{fan_role_2020}. Comparing the ratio of the widths of TAF and LSDW phases with that in the theoretical phase diagram in \cite{fan_role_2020}, we can estimate the degree of the interchain exchange anisotropy as $\epsilon = J_\perp^{xy}/J_\perp^z \approx 0.15$. The small value of this estimate supports our assumption of the dominant Ising-like nature of the interchain interaction.

The transition at $B^*$, evidenced in Fig.~\ref{fig:plateaux}c, and the exact nature of the phases remain not understood. We label them as TAF1 and TAF2 because the ordering vector detected in neutron scattering stays the same, see Fig.~\ref{figsm:001_vs_field}. 
One possibility is a change in the transverse moment orientation at this field, since $g_\perp$ along the c and b axes might be different.
The situation could also be similar to  \Bacovo\, which too features an unknown high-field state between the TAF and FP ones \cite{Klanjsek2015,Takayoshi2023}. We speculate that such an additional phase may be caused by the dipole-dipole interaction between spin chains.

\textit{Summary}--- Magnetization plateaux in spin-1/2 quantum magnets are rare and interesting. Their previous sightings include $M_s/3$ plateau state, also known as the {\em up-up-down} state, in spatially anisotropic triangular antiferromagnets Cs$_2$CuBr$_4$ \cite{fortune2009} and Cs$_2$CoBr$_4$ \cite{Facheris2022}. As explained above, magnetization plateaux are to be expected in quasi-one-dimensional magnets supporting the field-induced LSDW phase. Yet, 
Ising-like chain materials \Bacovo\ and \Srcovo\ that do feature LSDW 
phase do not appear to contain any finite-$M$ plateaux. We attribute this difference with YbAlO$_3$ magnet to the {\em antiferromagnetic} sign of the interchain interaction, which suppresses the {\em odd}-$k$ plateaux, in those Co-based magnets.

To the best of our knowledge, the reported observation of the {\em two} plateaux, at magnetizations $M_{2,5}$ and $M_{1,3}$, is the first of its kind. The fact that this is done with the help of heat transport measurement makes it even more rare. We hope that our findings generate further interest in unusual ordered states of quasi-one-dimensional magnetic materials.

\section{Acknowledgements}
We thank Cristian Batista, Leon Balents, Achim Rosch, Rong Yu, Jianda Wu, Burkhard Schmidt, Andy Mackenzie, and the participants of the KITP QMagnets23 program for fruitful discussions. E.H. acknowledges funding from Deutsche
Forschungsgemeinschaft (DFG) for the projects CRC 1143 - project number 247310070 (project C10) and the Wuerzburg-Dresden cluster of
excellence EXC 2147 "ct.qmat Complexity and Topology in Quantum Matter" - project number 390858490. Furthermore, E.H. and P.M. acknowledge support from the DFG for the CRC 80 - project number 107745057 (project E03) and from the Max Planck Society for the research group "Physics of Unconventional Metals and Superconductors" and the Max-Planck Fellowship.
O.A.S. acknowledges support by the NSF CMMT Grant No. DMR-1928919. This research
was supported in part by grant NSF PHY-2309135 to the Kavli Institute for Theoretical Physics (KITP).

\bibliographystyle{apsrev4-2}
%

\clearpage

\pagebreak
\widetext
\begin{center}
\textbf{\large Supplemental Materials}
\end{center}
\setcounter{equation}{0}
\setcounter{figure}{0}
\setcounter{table}{0}
\setcounter{page}{1}
\makeatletter
\renewcommand{\theequation}{S\arabic{equation}}
\renewcommand{\thefigure}{S\arabic{figure}}



\title{Supplementary Material}
\maketitle

\section{S1: Experimental Techniques}
Single crystals of \ybalo\ were prepared by a  Czochralski technique, as described elsewhere \cite{Buryy_growth_2010,noginov_role_2001}. For all measurements presented here, magnetic field is applied along the $a$ axis of the crystal.

The magnetization was measured using a high-resolution Faraday magnetometer \cite{sakakibara_magnetometer_1994} in a dilution cryostat. The data have been reproduced from \cite{wu_tomonagaluttinger_2019}.

\begin{figure}
    \center{\includegraphics[width=0.6\columnwidth]{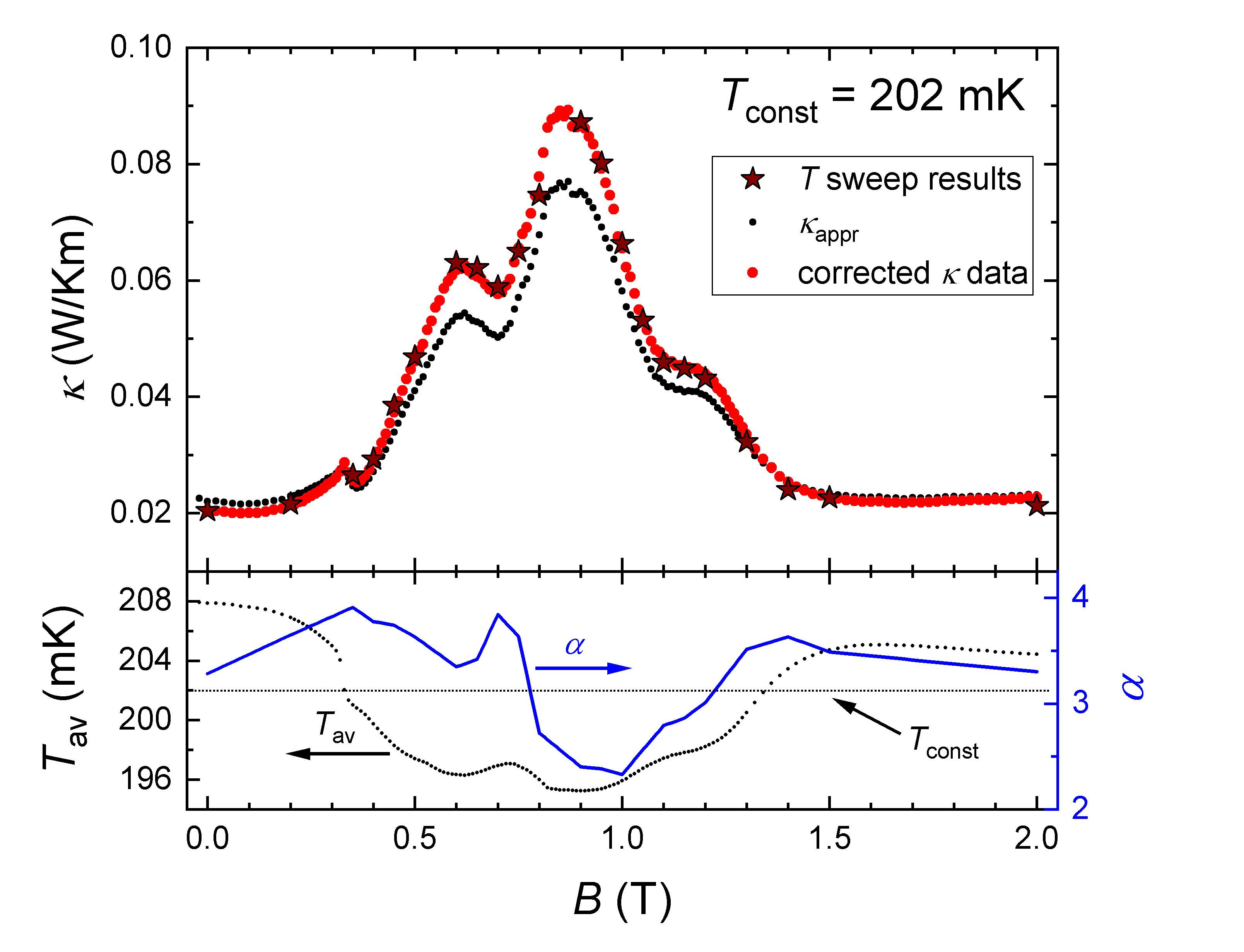}}
     \caption{a) Raw and corrected thermal conductivity data at $T_\mathrm{const} = 202$~mK. b) Variation of the average sample temperature during the field sweep (left axis) and the exponent $\alpha$ determined from temperature sweeps $\kappa (T) \propto T^\alpha$ around 202 mK (right axis). $T_\mathrm{const}$ used as a basis for calculation of the field sweep data is marked by a dotted line.}
     \label{fig:field-sweep-k}
\end{figure}

Thermal conductivity $\kappa (T) = \frac{P}{\Delta T}\frac{l}{A}$ is  obtained from measurements of the temperature gradient $\Delta T = T_\mathrm{w}-T_\mathrm{c}$ induced by a heat current through the sample generated by the heater power $P$, including the geometry factor $l/A$ of the sample where $l$ is the separation of the thermometer contacts and $A$ the cross section of the sample. This gives $\kappa$ at the average sample temperature $T_\mathrm{av} = (T_\mathrm{w}+T_\mathrm{c})/2$. We used a standard steady-state method with a two-thermometer-one-heater configuration in a \SP{3}He-\SP{4}He dilution refrigerator. The sample had a total length of $L=1.95$\,mm and a cross section $A = 0.5\times 0.5$\,mm$^2$, and the distance between the thermometer contacts was $l = 1.15$\,mm. The heat current was applied along the spin chains, i.e., the $c$ axis, and the temperature difference $\Delta T$ between the thermometer contacts was kept below 3\% of the average sample temperature.\\
During temperature sweeps, the sample thermometers were calibrated in-situ against a field-calibrated reference thermometer by measuring without heat current. This is important, because even tiny changes in the calibration curves well below 1~\% have large effects on the calculated temperature gradient, although they are not relevant for absolute temperatures.\\
For the magnetic field sweeps this in-situ calibration procedure is not applicable. In this case measurements were performed at constant bath temperature and gradient heater power. The direct calculation of the field- and temperature dependent temperature gradient $\Delta T$ and thermal conductivity $\kappa$ from the raw data is impeded by three problems: (1) the unknown sample thermometer calibration curves in zero field, (2) the field dependence of these calibrations, and (3) the change in average sample temperature $T_\mathrm{av}$ with magnetic field caused by the strong field-dependence of $\kappa$ and in consequence large variation of  $\Delta T$ at constant gradient heater power $P$. \\
To overcome the first problem (1), we started with an arbitrary calibration curve in zero field from a previous $T$ sweep and calculated approximate values $\Delta T_\mathrm{appr}$ and $\kappa _\mathrm{appr}$. $\kappa _\mathrm{appr}$ versus $B$ is shown in Fig. \ref{fig:field-sweep-k}  in comparison with the values of $\kappa$ from temperature sweeps. Owing to small deviations from the actual calibration curves, the calculated temperature difference $\Delta T_\mathrm{appr}$ contains an offset $\delta T$ corresponding to an apparent gradient at zero heat current, i.e., $\Delta T_\mathrm{appr} = \Delta T + \delta T$. $\delta T$ is assumed constant during field sweeps because changes of the calibration curves of the sample thermometers with time arise mainly from thermal cycling of the contacts and the setup during heating. However, as stated above, the average sample temperature remains basically unaffected by this tiny offset and is shown in the lower panel of Fig. \ref{fig:field-sweep-k}. Hence, the corrected $\kappa (T_\mathrm{av})$ thus far is then obtained from $\kappa (T_\mathrm{av}) = P/\Delta T = P/(\Delta T_\mathrm{appr}-\delta T) = 1/(1/\kappa _\mathrm{appr} - \delta T/P)$, where $\delta T$ is still unknown. \\
The second effect (2) is small in the field range of interest being below 3~\%  for $\kappa$ at $B \leq 2$~T. It is therefore ignored. \\
The last problem (3) concerns the varying average sample temperature $T_\mathrm{av}$ between the hot and cold thermometers, which - in combination with the strong temperature dependence of $\kappa$ - leads to considerable systematic errors in the field-sweep curves. When $\kappa$ strongly increases in intermediate fields near 0.8\,T (see Fig. \ref{fig:field-sweep-k}), the temperature gradient and average temperature decrease. Hence, the measured thermal conductivity corresponds to a lower value at lower temperature, so that $\kappa_\mathrm{appr}(T_\mathrm{av})$ stays below the values from the temperature sweeps. 
To get $\kappa (B,T_\mathrm{const})$ at constant $T_\mathrm{const}$, we applied a correction using a power law approximation of $\kappa (T) \propto T^\alpha$. This leads to the following correction to calculate the field-dependent real $\kappa$ at constant $T_\mathrm{const}$: 
\be
\kappa (B,T_\mathrm{const})= \frac{1}{1/\kappa_\mathrm{appr}(B,T_\mathrm{av}) - \delta T/P} \Big(\frac{T_\mathrm{const}}{T_\mathrm{av}(B)}\Big)^\alpha
\label{eqkappa}
\ee
The temperature- and field-dependent exponent $\alpha (B,T)$ was determined from the temperature sweeps and interpolated linearly between those fields as given in the lower panel of Fig.~\ref{fig:field-sweep-k}. $T_\mathrm{const}$ was taken as the average temperature  of maximum and minimum values of the $T_\mathrm{av}$ during a field sweep. The single free parameter $\delta T$ was adjusted to minimize the sum of quadratic deviations between results from $T$ and $B$ sweeps. Fig.~\ref{fig:field-sweep-k} compares $\kappa _\mathrm{appr}$ and $\kappa (B,T_\mathrm{const})$ for a field sweep at $T_\mathrm{const} = 202$~mK. The average temperature $T_\mathrm{av}$ during the field sweep varied between 196 mK and 208 mK. The difference between field-sweep and temperature-sweep results is below 6.5 \% for this curve, a value valid for 97 \% of all data points. Importantly, the features discussed in this manuscript appear in the raw data and are not altered by this data analysis.

\begin{figure}
    \center{\includegraphics[width=0.6\columnwidth]{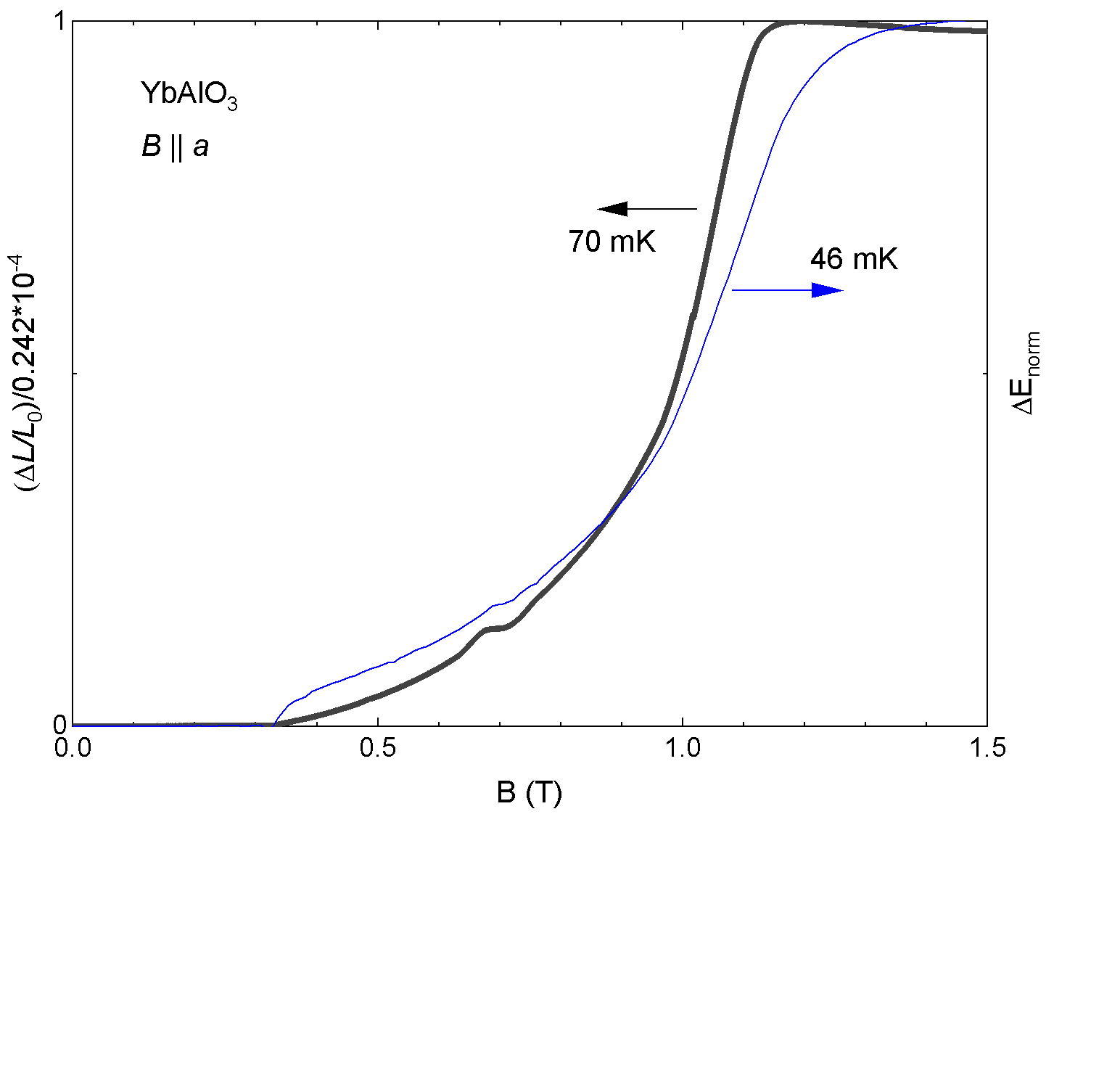}}
     \caption{Normalised $\Delta L/L_0$ at $T_\mathrm{const} = 70$~mK compared with the ``spin-interaction energy'' as defined in the text.}
     \label{fig:magnetostriction_raw}
\end{figure}

The magnetostriction measurements were performed in a dilution
refrigerator insert of a PPMS DynaCool using the world\`s smallest
high resolution capacitance dilatometer \cite{kuechler_new_applications_2023}. It was the first time that
magnetostriction measurements could be carried out in such a system at
low temperatures in between 70 mK and 4 K \cite{kuechler_smallest_2017}. The magnetostriction
was measured on a YbAlO3 single crystal along the $c$ axis with a length of
$L_0$ = 1.74 mm. Here, the change in the length along c, $\Delta L$ was measured while
the magnetic field was applied perpendicularly, i.e., along $a$.
The normalised magnetostriction data $\Delta L/L_0$ is shown in Fig.~\ref{fig:magnetostriction_raw} and compared with the normalised ``spin-interaction energy'' $\Delta E = \int_0^M B(M') dM'$. The similar behavior is evidence that the magnetostriction can be dominantly understood as exchange-striction \cite{Miyata2021}.
In order to obtain the magnetostriction coefficient $\lambda = \frac{1}{L_0} \dv{L}{B}$ the raw data is derived with respect to the magnetic field with a mild Savitzky-Golay smooting routine to reduce the noise in the derivative. The sweep-field rate of the 70\,mK curve was much smaller than that of the higher temperature curves. This is why the transition at $B^*$ is only sharp in that curve and only this temperature is shown in Fig.~3f of the main paper.

\section{S2: Temperature range of the plateau states}

For both plateau states, the anomalies in the magnetostriction are observed up to 250\,mK but disappear at 300\,mK. For the 1/5 plateau $T_\mathrm{N}(0.5\,T)= 300$\,mK but for the 1/3 plateau $T_\mathrm{N}(0.7\,T)= 240$\,mK, so that the plateau signature seems to extend above $T_N$ for the latter. However, $T_N$ as shown in our phase diagram was previously defined at the peak of the specific heat anomaly. The anomaly has a width of typically 40\,mK, so that the bulk phase transition is more close to 260 mK. This means that the anomaly is only present below $T_\mathrm{N}$ as confirmed by the thermal conductivity in Fig~\ref{fig:temp_evol} at 259\,mK for which the anomaly is absent.

\section{S3: Estimation of the gap in the 1/3 plateau state}
\label{sm-1/3gap}
\begin{figure*}[!ht]
     \includegraphics[width=0.8\textwidth]{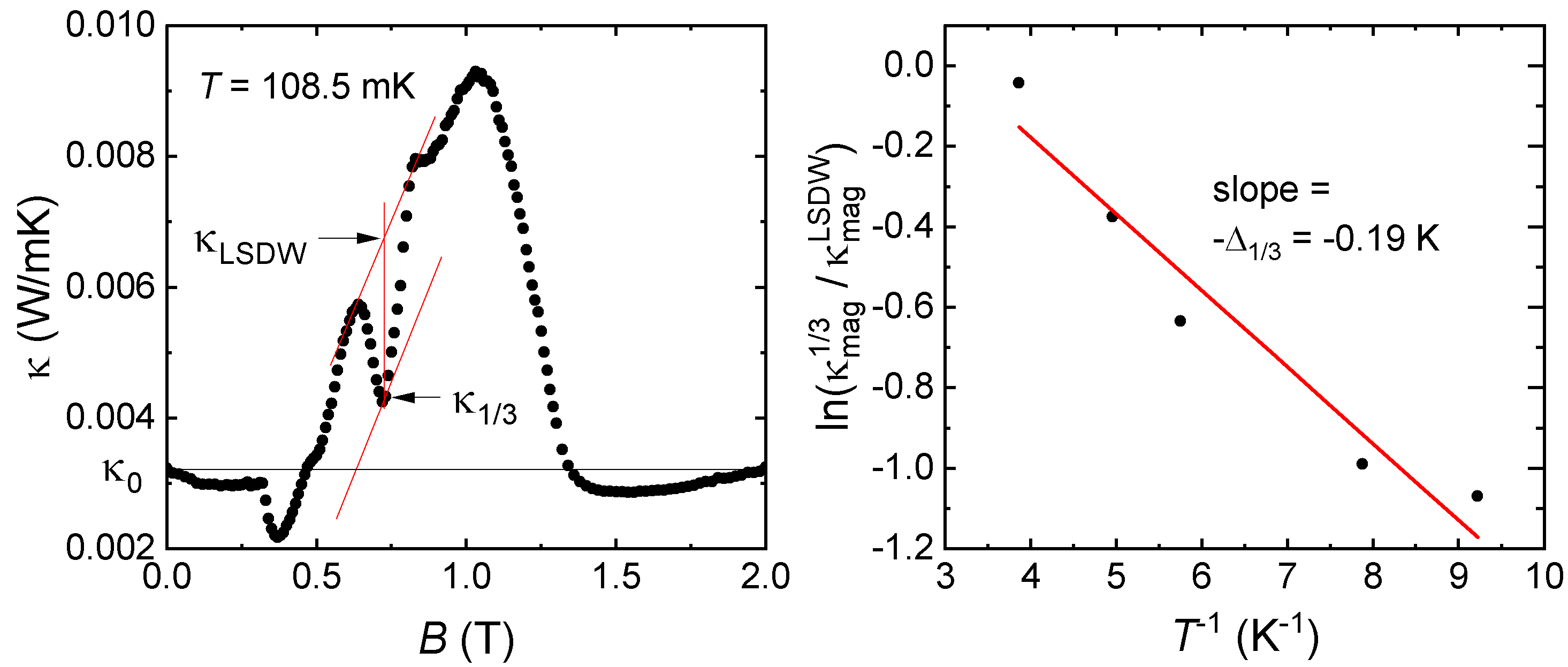}
   \caption{
    a) Raw data of thermal conductivity $\kappa(B)$ at 108.5\,mK with an extraction of characteristic values of $\kappa_\mathrm{1/3}$, $\kappa_\mathrm{LSDW}$, and $\kappa_{0}$. b) Estimation of the gap in the 1/3 plateau state $\Delta_{1/3}$.
    }
    \label{figsm:kapparaw}
\end{figure*}

Fig.\ref{figsm:kapparaw}a shows the thermal conductivity data for 108.5\,mK as an example. $\kappa_0$ is the phonon contribution when phonons cannot scatter with magnetic excitations since those are gapped for both low and high fields. For each temperature we estimate the magnetic contribution to heat transport as $\kappa_\mathrm{mag} = \kappa-\kappa_0$. This represents a lower limit of $\kappa_\mathrm{mag}$ because additional scattering of phonons with magnetic excitations in the intermediate field range might reduce the phonon contribution to thermal transport and would lead to a larger magnetic contribution. In order to get a value of the gap, we assume that the temperature dependence inside the gapped phase of $\kappa_\mathrm{mag}^{1/3} = \kappa_{1/3}-\kappa_0$ can be roughly written as $\kappa_\mathrm{mag}^{1/3}=\kappa_\mathrm{mag}^\mathrm{LSDW}\cdot\exp{(-\Delta/k_\mathrm{b}T)}$. Here $\kappa_\mathrm{mag}^\mathrm{LSDW} = \kappa_\mathrm{LSDW}-\kappa_0$ is an estimate for the thermal conductivity in the LSDW state at the same field of the plateau. Although the temperature dependence does not follow a perfect exponential function, this analysis also gives a gap of 0.19\,K as shown in panel b.

\section{S4: Theoretical description of magnetization plateaux}
The theoretical analysis below closely follows Section III.D of Ref.\cite{extreme} where the detailed theoretical description of magnetization plateaux in quasi-one-dimensional magnets was developed. Some details are also described in Ref.\cite{nematic}. This and related physics are extensively reviewed in \cite{Starykh2015}.

The key idea is that the magnetization plateau is the commensurate version of the longitudinal spin-density wave (SDW) state. Longitudinal SDW is the state where spins are ordered along the direction of the applied magnetic field. Quite generally, this is an incommensurate state with the ordering momentum given by $\pi \pm 2\delta$, where $\delta = \pi M$ and $M$ is the magnetization. Commensurate magnetization plateau is made possible by {\em umklapp} processes when the quasi-momentum of $k$ SDW excitations (thus $k$ is an {\em integer} number) is equal to $\pm 2\pi$ and thereby can be emitted or absorbed by the lattice as a whole (this is the process that distinguished lattice quasi-momentum from free-space momentum). This corresponds to the locking-in of the incommensurate SDW state into a commensurate plateau state. 


We assume the following sequence of energy scales: chain exchange interaction $J$ is the largest one, the interchain interaction $J'$ is the next one, and the umklapp interaction $t_k$ is the smallest. Thus, $J \gg J' \gg t_k$. 

\subsection{Symmetry imposed constraints}

We start with Eq.32 of Ref.\cite{extreme} for the longitudinal component $S^z_y(x)$ of the spin operator at the site $x$ on the chain with index $y$. Both $x$ and $y$ are integers, measured in units of corresponding lattice spacings (along the chain and between the chains, correspondingly) which we set to unity. Then in the {\em continuum} limit 
\be
S^z_y(x) \sim M + \frac{1}{\beta} \partial_x \phi_y(x) - A_1 \sin\Big[\frac{2\pi}{\beta} \phi_y(x) - (\pi - 2\delta)x\Big] .
\label{eq1}
\ee
Here $\beta$ is the parameter of the theory and is eventually related to the Luttinger parameters of the one-dimensional field theory.  Eq.\eqref{eq1} shows that field $\phi_y(x)$ is periodic with period $\beta$, so that $\phi_y(x) \to \phi_y(x) + \beta$ must always hold.

Observe that $\phi_y(x)$ describes quantum fluctuation of the state with the fixed magnetization $M$ per chain site: $\langle S^z_y(x)\rangle = M + \frac{1}{\beta} \partial_x \langle \phi_y(x) \rangle = M$. This means that $\langle \phi_y(x) \rangle$ is $x$-independent (and angular brackets denote the expectation value with respect to the fixed magnetization state). 

Now imagine {\em actively translating} the field $S^z_y(x)$ along the chain by exactly one lattice spacing. Noting that $\partial_x \phi_y(x+1) \approx \partial_x \phi_y(x)$ upto higher derivatives, we obtain
\be
S^z_y(x+1) \sim M + \frac{1}{\beta} \partial_x \phi_y(x) - A_1 \sin\Big[\frac{2\pi}{\beta} \phi_y(x+1) - (\pi - 2\delta)(x+1)\Big]
\label{eq2}
\ee
and conclude that under the translation along the chain continuum field $\phi_y(x)$ transforms as 
\be
\phi_y(x) \to \phi_y(x+1) - \frac{\beta}{2\pi} (\pi - 2\delta) .
\label{eq3}
\ee
Similarly, spatial inversion of the field $S^z_y(x)$ results in 
\be
\phi_y(x) \to \frac{\beta}{2} - \phi_y(-x).
\label{eq4}
\ee
In the rectangular geometry appropriate for YbAlO$_3$ translating $S^z_y(x)$ from chain $y$ to the neighboring chain $y+1$ only changes the discrete index $y$
\be
\phi_y(x) \to \phi_{y+1}(x) .
\label{eq5}
\ee
and does not play an important role in the following. (This is different from the triangular geometry where the translation along the diagonal leads to Eq.34 in Ref.\cite{extreme}.) 

We now write down the $k$-th order umklapp term and use the above symmetries to constrain its form, 
\be
H^{(k)}_{\rm umk} = \sum_y \int dx \, t_k \, \cos\Big[\frac{2\pi k}{\beta} \phi_y(x) + \zeta_k\Big]
\label{eq6}
\ee
where $t_k \sim o(J)$ is the amplitude and $\zeta_k$ is a $k$-dependent phase factor. Periodicity $\phi_y(x) \to \phi_y(x) + \beta$ requires $k$ to be integer.

Under translation \eqref{eq3} the argument of cosine in \eqref{eq6} changes like
\be
\frac{2\pi k}{\beta} \phi_y(x) + \zeta_k \to \frac{2\pi k}{\beta} \Big[\phi_y(x+1)  - \frac{\beta}{2\pi} (\pi - 2\delta)\Big] + \zeta_k = \frac{2\pi k}{\beta} \phi_y(x+1) + \zeta_k - k (\pi - 2\delta)
\label{eq7}
\ee
which requires 
\be
k (\pi - 2 \delta) \equiv k \pi (1-2M) = 2\pi \nu
\label{eq8}
\ee
where $\nu$ is another integer. Therefore 
\be
M = M_{\nu, k} = \frac{1}{2}\Big(1-\frac{2\nu}{k}\Big)
\label{eq9}
\ee
defines the set of magnetizations $M_{\nu, k}$ for which $k$-th order umklapp interaction \eqref{eq6} is allowed in the Hamiltonian. Observe that $M \geq 0$ means that $\nu \leq k/2$.

Under inversion \eqref{eq4} 
\be
\frac{2\pi k}{\beta} \phi_y(x) + \zeta_k \to \frac{2\pi k}{\beta}\Big(\frac{\beta}{2} - \phi_y(-x)\Big) + \zeta_k = - \frac{2\pi k}{\beta} \phi_y(-x) + \zeta_k + \pi k
\label{eq10}
\ee
Since cosine is even function of its argument this means $\zeta_k = - \zeta_k - \pi k$ and thus the phase in \eqref{eq6} is
\be
\zeta_k = - \frac{\pi}{2} k
\label{eq11}
\ee

\subsection{Energetic considerations}

Our next goal is to minimize the in-chain umklapp interaction \eqref{eq6} simultaneously with the inter-chain SDW term (which is responsible for the SDW state in the first place). The relevant interchain interaction for XXZ chains is discussed in Section II.B.1 of Ref. \cite{nematic}, it is given by the first term of Eq.(12) of that reference 
\be
H_{\rm sdw} = \sum_y \int dx \, \gamma_{\rm sdw} \cos[\frac{2\pi}{\beta} (\phi_y(x) - \phi_{y+1}(x))]
\label{eq12}
\ee
where the coupling constant is estimated to be $\gamma_{\rm sdw} = J' \Delta A_1^2/2$, see Table I in \cite{nematic}.  Here $J'$ is the interchain exchange interaction and $\Delta$ is in-chain XXZ anisotropy (note notational confusion: in \cite{nematic} the XXZ anisotropy is denoted by $\delta$). 

Configuration of the field $\phi_y(x)$ that minimizes \eqref{eq12} depends on the sign of $\gamma_{\rm sdw} \sim J'$.

{\bf 1. $J' < 0$.} Let $J'$ to be ferromagnetic so that $\gamma_{\rm sdw} = - |J'| \Delta A_1^2/2 < 0$ is negative. We see that $\phi_y(x) - \phi_{y+1}(x) = \beta \ell$, with integer $\ell$, trivially minimizes \eqref{eq12}. But since $\phi_y(x)$ is periodic in $\beta$ anyway, the classical equilibrium configuration is given by simple $\phi_y(x) = \phi_{y+1}(x)$ condition. More accurately, we split the field into its classical part $\phi_y^{(0)}$ and quantum fluctuations $\tilde{\phi}_y(x)$ on top of it,
\be
\phi_y(x) \to \phi_y^{(0)} + \tilde{\phi}_y(x)
\label{eq13}
\ee
Then our brief analysis above shows that Eq.\eqref{eq12} is minimized by $\phi_y^{(0)} = \beta \ell$ which is equivalent to $\phi_y^{(0)} = 0$.

This observation does not affect the form $H_{\rm umk}^{(k)}$ in \eqref{eq6}, \eqref{eq8} and \eqref{eq11}, and therefore does not produce any additional constraints on integers $\nu$ and $k$ determining the allowed magnetization values of the plateaux in \eqref{eq9}.

{\bf 2. $J' > 0$.} Let now $J'$ to be antiferromagnetic so that $\gamma_{\rm sdw} > 0$. This requires 
\be
\phi_y^{(0)} - \phi_{y+1}^{(0)} = \frac{\beta}{2} (2\ell +1)
\label{eq14}
\ee
This is satisfied by 
\be
\phi_y(x) \to \frac{\beta}{2} y  + \tilde{\phi}_y(x)
\label{eq15}
\ee
which leads to a change of the overall sign of $H_{\rm sdw}$ in \eqref{eq12} when expresses in terms of ``new" (fluctuating) field $\tilde{\phi}_y(x)$.

This substitution changes the umklapp Hamiltonian to
\be
H^{(k)}_{\rm umk} = \sum_y \int dx \, t_k \, \cos\Big[\frac{2\pi k}{\beta} \tilde{\phi}_y(x) + \pi k y - \frac{\pi}{2} k \Big]
\label{eq16}
\ee
For {\bf even} $k = 2n$ the term  $\pi k y = 2 \pi n y$ is a multiple of $2\pi$ for any chain index $y$ and therefore drops out of \eqref{eq16}. Then
\be
H^{(k=2 n)}_{\rm umk} = \sum_y \int dx \, t_k \, \cos\Big[\frac{4\pi n}{\beta} \tilde{\phi}_y(x) - \pi n \Big]
\label{eq17}
\ee
with $y$-independent cosine. This means that $H^{(k=2 n)}_{\rm umk}$ is minimized the same way in every chain $y$. The total energy gain due to the umklapp interaction of $k=2n$ order is given by that in the single chain times the (infinite) number of chains, i.e. it is extensive.

The situation is different for {\bf odd} $k = 2 n + 1$. Now $\pi k y = \pi (2 n+1) y = 2\pi n y + \pi y \to \pi y$ (since $2\pi y$ is multiple of $2\pi$ for every integer $y$). Thus 
\bea
&&H^{(k=2n+1)}_{\rm umk} =  \int dx \, t_k \, \sum_y \cos\Big[\frac{2\pi (2n+1)}{\beta} \tilde{\phi}_y(x) + \pi y - \frac{\pi}{2} (2n+1) \Big] =\nonumber\\
&& \int dx \, t_k \, \Big\{ \sum_{y = {\rm even}} \cos\Big[\frac{2\pi (2n+1)}{\beta} \tilde{\phi}_y(x) - \frac{\pi}{2} (2n+1) \Big] - \sum_{y = {\rm odd}} \cos\Big[\frac{2\pi (2n+1)}{\beta} \tilde{\phi}_y(x) - \frac{\pi}{2} (2n+1) \Big] \Big\}
\label{eq18}
\eea
where we split the $y$-sum into that over $y = {\rm even}$ and $y = {\rm odd}$ subsets and used the basic fact that $\cos[\alpha + \pi \cdot {\rm odd~integer}] = - \cos[\alpha]$. We see that the expectation value of \eqref{eq18} is zero, and the contributions of {\em even}-numbered and {\em odd}-numbered subsets of chains cancel each other exactly.

That is, magnetization plateaux of the {\em odd} order ($k=2n+1$) are energetically disfavored for $J' > 0$.

\subsection{Experimental consequences}

Eq.\eqref{eq9} tells us the first member in the $M_{\nu, k}$ sequence of plateaux is given by the $1/3$-plateau $M_{1,3} = 1/2 \times (1 - 2/3) = 1/2 \times 1/3$ while $M_{2,5} = 1/2 \times (1 - 2 \times 2/5) = 1/2 \times 1/5$ corresponds to $\nu = 2, k = 5$. Notice that another ``natural" member $M_{1,5} = 1/2 \times (1 - 2/5) = 1/2 \times 3/5$ corresponds to the magnetization bigger than that of $M_{1,3}$. Note that $k$ is {\em odd} for both $M_{1,3}$ and $M_{2,5}$.

To obtain these plateaux with {\em even} $k$, we should consider, for example, $\nu = 2, k = 6$ state which gives $M_{2,6} = 1/2 \times (1 - 2 \times 2/6) = 1/2 \times 1/3$. So, magnetization-wise this state is equivalent to $M_{1,3}$ one because $M_{1,3} = M_{2,6}$. However, it turns out that higher values of $k$ result in much more narrow plateaux. Specifically, the width of the plateau is exponentially sensitive to the index $k$, see \eqref{eq39} below. ( The full renormalization-group-based argument is in Ref.\cite{extreme}.)

This means that plateaux with the smallest $k$ are the most stable and, therefore, most pronounced. 

Our analysis above shows that rectangular geometry with $J' < 0$ strongly favors odd-valued plateaux with odd $k$. In the case of $J' > 0$ odd $k$'s are energetically not possible.

Since YbAlO$_3$ is characterized by the ferromagnetic inter-chain exchange, $J' < 0$, we conclude that plateaux at $1/3$ of the total magnetization ($M_{1,3}$) and at $1/5$ of it ($M_{2,5}$) are the two most pronounced of their kind, with other plateaux either having higher $k$ (and therefore being more narrow in field) or occurring at higher magnetization where the SDW state ceases to be the ground state. It thus appears that our theoretical description is consistent with the experiment and, in fact, explains what makes YbAlO$_3$ special.

Perhaps a little Table of possible $M_{\nu, k}$ will be useful here:
\begin{table}[h]
\begin{tabular}{lllllllllllll}
 $k$ &  3 & 4 & 5 & 5 &  6 & 6 & 7 & 7 & 7 & 8 & 8 & 8 \\
 $\nu$ & 1 & 1  & 1 & 2 & 1 & 2 & 1 & 2 & 3 & 1 & 2 & 3\\
 $2 M$ & {\color{red} $\frac{1}{3}$} & $\frac{1}{2}$ & $\frac{3}{5}$  & {\color{red}$\frac{1}{5}$} & $\frac{2}{3}$ & $\frac{1}{3}$ & $\frac{5}{7}$ & $\frac{3}{7}$ & $\frac{1}{7}$ & $\frac{3}{4}$ & $\frac{1}{2}$ & $\frac{1}{4}$
\end{tabular}
\end{table}

\subsection{Plateaux}

We now focus on the relevant case of ferromagnetic interchain exchange so that $\gamma_{\rm sdw} < 0$.

In the absence of the chain umklapp term \eqref{eq6} magnetization varies with the magnetic field monotonously and commensurate magnetizations $M_{\nu, k}$ constitute set of measure zero. That is, magnetization $M(h)$ takes particular value $M_{\nu, k}$ at a particular (and unique) value of the field $h=h_{\nu, k}$ such that $M(h_{\nu, k}) = M_{\nu, k}$. $k$-the order magnetization plateau appears when 
\be
H^{(k)}_{\rm umk} = \sum_y \int dx \, t_k \, \cos\Big[\frac{2\pi k}{\beta} \phi_y(x) - \frac{\pi}{2} k\Big]
\label{eq20}
\ee
is strong enough to pin down $\phi_y(x)$ at a particular value minimizing \eqref{eq20} in a {\em finite interval} of field $h$ in the neighborhood $h_{\nu, k}$. To describe this quantitatively we need to generalize our description to $h$ near but not equal to $h_{\nu, k}$. 

First, we need to bring \eqref{eq20} into a minimization-ready form. Given discussion of $M_{\nu, k}$ values in the Table above, we focus on the case of {\em odd} $k$ (extension to even $k$ is obvious). Then 
\be
H^{(k)}_{\rm umk} = \sum_y \int dx \, t_k (-1)^{\frac{k-1}{2}} \, \cos\Big[\frac{2\pi k}{\beta} \phi_y(x) - \frac{\pi}{2} \Big]
\label{eq21}
\ee
Classical configuration $\phi^{(0)}$ (see \eqref{eq13}) that minimizes \eqref{eq21} is determined by the sign of $\tilde{t}_k = t_k (-1)^{\frac{k-1}{2}}$. 

For $t_k (-1)^{\frac{k-1}{2}} < 0$ we need $\frac{2\pi k}{\beta} \phi^{(0)} = \frac{\pi}{2}$ and therefore we write ($\Theta$ is the Heavyside step function)
\be
\phi_y(x) = \frac{\beta}{4 k} \Theta(-\tilde{t}_k) + \tilde{\phi}_y(x) .
\label{eq22}
\ee
For $t_k (-1)^{\frac{k-1}{2}} > 0$ we need $\frac{2\pi k}{\beta} \phi^{(0)} = -\frac{\pi}{2}$ and therefore
\be
\phi_y(x) = -\frac{\beta}{4 k} \Theta(\tilde{t}_k) + \tilde{\phi}_y(x) .
\label{eq23}
\ee
In both cases $\phi_y(x) = -\frac{\beta}{4 k} {\rm sgn}(\tilde{t}_k) + \tilde{\phi}_y(x)$ and we end up with 
\be
H^{(k)}_{\rm umk} = \sum_y \int dx \, \Big( - |t_k (-1)^{\frac{k-1}{2}}|\Big) \, \cos\Big[\frac{2\pi k}{\beta} \tilde{\phi}_y(x) \Big]
\label{eq24}
\ee
which is written in terms of fluctuating component $\tilde{\phi}_y(x)$ of the original field $\phi_y(x)$. Importantly, constant shifts in \eqref{eq22} and \eqref{eq23} do not affect spatial derivatives of the field. That is, $\partial_x \phi_y(x) = \partial_x \tilde{\phi}_y(x)$. Being $y$-independent, they also do not affect the interchain SDW interaction \eqref{eq12}, 
$\phi_y(x) - \phi_{y+1}(x) = \tilde{\phi}_y(x) - \tilde{\phi}_{y+1}(x)$.

According to Eq.\eqref{eq1} magnetization per site is 
\be
\langle S^z_y(x) \rangle = M + \frac{1}{\beta} \partial_x \langle \tilde{\phi}_y(x) \rangle .
\label{eq25}
\ee

Now, we claim that magnetic field $h$ in a small neighborhood of $h_{\nu, k}$ is described by the following modification \cite{LB2008} of \eqref{eq24}
\be
H^{(k)}_{\rm umk} = \sum_y \int dx \, \Big( - |t_k (-1)^{\frac{k-1}{2}}|\Big) \, \cos\Big[\frac{2\pi k}{\beta} \tilde{\phi}_y(x) + \alpha x\Big]
\label{eq26}
\ee
Being the only field-dependent term in the full Hamiltonian, \eqref{eq26} is minimized by 
\be
\tilde{\phi}_y(x) = - \frac{\alpha \beta}{2\pi k} x + \varphi_y(x)
\label{eq:o1}
\ee
where
$\langle \varphi_y(x) \rangle =0$. Then \eqref{eq25} becomes
\be
\langle S^z_y(x) \rangle = M - \frac{1}{\beta} \partial_x \Big(\frac{\alpha \beta}{2\pi k} x\Big) = M - \frac{\alpha}{2\pi k}
\label{eq27}
\ee
We now {\em require} that $\langle S^z_y(x) \rangle = M_{\nu, k}$ (that is, that magnetization $\langle S^z_y(x) \rangle$ does not depend on field $h$ and therefore exhibits the plateau behavior) which immediately fixes $\alpha$ to be 
\be
\alpha(h) = 2\pi k \Big( M(h) - M_{\nu, k}\Big)
\label{eq28}
\ee
Observe that by construction $\alpha(h_{\nu, k}) = 0$. Thus, $\alpha(h)$ in \eqref{eq28} is the wavevector parameterizing variation of the magnetic field in the vicinity of the `optimal' $h_{\nu, k}$ value. Keep in mind that $M(h)$ here represents magnetization per site in the {\em absence} of the umklapp interaction, while $\langle S^z_y(x) \rangle$ is that of the {\em full} theory.

To connect with Ref.\onlinecite{extreme} the Hamiltonian of the full system can be written in terms of $\varphi$ field
\bea
H^{(k)}_{\rm plateau} &=& \sum_y \int dx \Big\{\frac{v}{2} (\partial_x \varphi_y(x))^2 + \frac{v}{2} (\partial_x \theta_y(x))^2 - |\gamma_{\rm sdw}| \cos\Big[\frac{2\pi}{\beta} (\varphi_y(x) - \varphi_{y+1}(x))\Big]   \nonumber\\
&& - \frac{v \beta \alpha}{2\pi k} \partial_x \varphi_y(x) - |t_k (-1)^{\frac{k-1}{2}}| \cos\Big[\frac{2\pi k}{\beta} \varphi_y(x)\Big] \Big\}
\label{eq29}
\eea
which is just Eq.48 of \cite{extreme} (upto what looks like misprints in the linear derivative term there). The same Hamiltonian can also be written in terms of the `original' field \eqref{eq:o1} $\tilde{\phi}_y(x)$ as
\bea
H^{(k)}_{\rm plateau} &=& \sum_y \int dx \Big\{\frac{v}{2} (\partial_x \tilde{\phi}_y(x))^2 + \frac{v}{2} (\partial_x \theta_y(x))^2 - |\gamma_{\rm sdw}| \cos\Big[\frac{2\pi}{\beta} (\tilde{\phi}_y(x) - \tilde{\phi}_{y+1}(x))\Big]  \nonumber\\
&& - |t_k (-1)^{\frac{k-1}{2}}| \cos\Big[\frac{2\pi k}{\beta} \tilde{\phi}_y(x) + \alpha x\Big]\Big\}
\label{eq30}
\eea
In both Hamiltonians field $\theta_y(x)$ is the field dual to $\phi_y(x)$ (and, therefore, also dual to $\tilde{\phi}_y(x)$ and $\varphi_y(x)$).  The dual field is crucial for understanding the dynamics of $\phi_y(x)$. 

\subsection{Transition to the three/two-dimensional regime}

Since interchain SDW interaction is relevant, the ground state of our system is ordered in the longitudinal SDW phase. To describe it, we follow steps outlined in \cite{extreme, LB2008}. The key idea is to integrate out high-energy modes in the interval $\Lambda_{\rm sdw} < k < \Lambda_0 \sim \pi/a_0$. Here $k = (\omega_n, k_x)$ is two-dimensional Euclidian momentum (and for convenience we have rescaled euclidian time by the factor of velocity, $\tau \to v \tau = x_0$, so that it has the same dimension as the coordinate along the chain), and $a_0 = a_x$ is the distance between nearest spins along the chain. Notice that chain index $y$ (in the three-dimensional system that index has two components, $y \to (y,z)$) remains discrete for now (it will become continuous coordinate ${\bm \rho} = (y, z) a_y$ at the end of the day, and $a_y$ is the minimal distance between neighboring chains). 

Now we can follow steps outlined in Section III.D.4 of \cite{extreme} to find that perturbative elimination of high-energy degrees results in the renormalization of the SDW coupling in \eqref{eq30} 
$\gamma_{\rm sdw}(0) \to \gamma_{\rm sdw}(\ell) = \gamma_{\rm sdw}(0) (\Lambda_{\rm sdw}/\Lambda_0)^2$. In the differential form ($\Lambda_{\rm sdw} = \Lambda_0 e^{-\ell}$) this is just
\be
\frac{\partial \gamma_{\rm sdw}(\ell)}{d\ell} = - 2 \Delta_{\rm sdw} \gamma_{\rm sdw}(\ell) \, \, \Rightarrow \gamma_{\rm sdw}(\ell) = \gamma_0 \Big(\frac{\Lambda_{\rm sdw}}{\Lambda_0} \Big)^{2\Delta_{\rm sdw}}
\label{eq31}
\ee
where $\Delta_{\rm sdw} = \pi/\beta^2$ is the scaling dimension of $\phi$ field and we introduced $\gamma_0 = \gamma_{\rm sdw}(\ell =0)$ for brevity. We fix the new momentum cut-off $\Lambda_{\rm sdw}$ by requiring the maximum energy along the chain $\sim v \Lambda_{\rm sdw}^2$ to be comparable to the renormalized SDW interaction between the chains $\gamma_{\rm sdw}(\ell)$ so that 
\be
\Lambda_{\rm sdw}^2 = \frac{\gamma_0}{v} \Big(\frac{\Lambda_{\rm sdw}}{\Lambda_0} \Big)^{2\Delta_{\rm sdw}} \, \, \Rightarrow \frac{\Lambda_{\rm sdw}}{\Lambda_0} = \Big( \frac{\gamma_0}{v \Lambda_0^2}\Big)^{\frac{1}{2(1-\Delta_{\rm sdw})}}
\label{eq32}
\ee
Observe that $\tilde{\gamma}_0 \equiv \gamma_0/(v \Lambda_0^2) \ll 1$ is small dimensionless parameter of the theory.

Thus, by construction, at the scale $\Lambda_{\rm sdw}$ the system is approximately three-dimensional and SDW cosine in \eqref{eq29} can be approximated as 
\be
\cos\Big[\frac{2\pi}{\beta} (\tilde{\phi}_y(x) - \tilde{\phi}_{y+1}(x))\Big] \approx 1 - (\frac{2\pi}{\beta})^2 a_y^2 [\partial_y \tilde{\phi}(x,y)]^2
\label{eq33}
\ee

The umklapp interaction renormalizes as well
\be
t_k \to t_k(\ell) = t_k(0) \Big(\frac{\Lambda_{\rm sdw}}{\Lambda_0} \Big)^{k^2  \Delta_{\rm sdw}}
\label{eq34}
\ee
where $t_k(0)$ is the initial umklapp coupling in \eqref{eq29} or \eqref{eq30}. Notice that the exponent $k^2  \Delta_{\rm sdw}$ grows very fast with the order $k$ of the umklapp process. Since $\Lambda_{\rm sdw}/\Lambda_0 < 1$, hence plateaux with large $k$ are strongly suppressed.

As a result, \eqref{eq30} can now be written in terms of the continuous 3-dimensional field $\phi(x,y)$ (here $y$ is $(y,z)$ as mentioned above and we drop {\em tilde} symbol $\tilde{}$ on top of $\phi$ so that $\tilde{\phi} \to \phi$) 
\be
H_{3d} \approx \int dx dy \Big\{ \frac{v}{2} (\partial_x \theta_y(x))^2 + \frac{v}{2} (\partial_x \phi(x,y))^2 + \frac{c_y}{2} (\partial_y \phi(x,y))^2 - |t_k(\ell) (-1)^{\frac{k-1}{2}}| 
\cos\Big[\frac{2\pi k}{\beta} \phi(x,y) + \alpha x\Big] \Big\}
\label{eq36}
\ee
where 
\be
c_y = v \, \big(\frac{2\pi}{\beta}\big)^2 (a_y \Lambda_0)^2   \Big( \frac{\gamma_0}{v \Lambda_0^2}\Big)^{\frac{1}{1-\Delta_{\rm sdw}}} \ll v
\label{eq37}
\ee
is the velocity in the transverse to the chain direction(s) and the coefficient of cosine is given by \eqref{eq34}.

We are now in position to estimate the width of the plateau by considering classical limit of \eqref{eq36}, i.e. by treating $\langle \phi(x,y)\rangle$ as classical, time-independent field. Cosine term in \eqref{eq36} is minimized by $\langle \phi(x,y)\rangle = - \alpha \beta x/(2\pi k)$ and the energy gain is $-|t_k(\ell)|$. However this $x$-dependent expectation value costs deformation (kinetic) energy 
$v (\alpha \beta/(2\pi k))^2/2$. Equating the two we find the critical $\alpha_c$,
\be
\alpha_c = \frac{2\pi k}{\beta} \sqrt{\frac{2 |t_k(0)|}{v}} \Big(\frac{\Lambda_{\rm sdw}}{\Lambda_0} \Big)^{k^2  \Delta_{\rm sdw}/2}
\label{eq38}
\ee
For $\alpha \leq \alpha_c$ the plateau is stable, while for $\alpha \geq \alpha_c$ it is replaced by the incommensurate SDW phase. Given \eqref{eq28} and the fact that in the narrow range of the plateau phase $dM/dh$ is constant (remember that here $M$ is the magnetization of the spin chain in the absence of the umklapp interaction term), we can use \eqref{eq38} to estimate the width of the plateau in magnetic field as
\be
\Delta h = \frac{1}{\beta dM/dh} \sqrt{\frac{2 |t_k(0)|}{v}} \Big(\frac{\Lambda_{\rm sdw}}{\Lambda_0} \Big)^{k^2  \Delta_{\rm sdw}/2} \sim 
 \Big( \frac{\gamma_0}{v \Lambda_0^2}\Big)^{\frac{k^2 \Delta_{\rm sdw}}{4(1-\Delta_{\rm sdw})}}
\label{eq39}
\ee
where the last estimate follows from \eqref{eq32}. {\em Very} approximately $\Delta_{\rm sdw} \approx 1/2$ so that the exponent of the last term in \eqref{eq39} is estimated as $k^2/4$. That is, $\Delta h \sim (J'/J)^{k^2/4}$. As already discussed previously: the higher the value of $k$, the more narrow the plateau is.

\subsection{Excitations of the SDW and plateau phases}

Finally we can address the issue of excitations described by the Hamiltonian \eqref{eq36}. Cosine term there makes the problem unsolvable but there are two easy and physically important limits that we can analyze. First is the incommensurate SDW phase which is obtained by eliminating the cosine term from \eqref{eq36}. The second limit is represented by the center of the magnetization plateau, when $\alpha=0$. In that limit we can expand the umklapp cosine
\be
\cos\Big[\frac{2\pi k}{\beta} \phi(x,y)\Big] \approx 1 - \frac{1}{2} \big(\frac{2\pi k}{\beta}\big)^2  \phi(x,y)^2 ,
\label{eq35}
\ee
so that \eqref{eq36} can be further approximated as 
\be
H_{3d} \approx \int dx dy \Big\{ \frac{v}{2} (\partial_x \theta(x,y))^2 + \frac{v}{2} (\partial_x \phi(x,y))^2 + \frac{c_y}{2} (\partial_y \phi(x,y))^2 + \frac{m_k^2}{2} \phi^2(x,y) \Big\}
\label{eq40}
\ee
where the mass $m_k$ is given by 
\be
m_k^2 = \big(\frac{2\pi k}{\beta}\big)^2  |t_k(\ell)| .
\label{eq41}
\ee
So obtained $H_{3d}$ in \eqref{eq40} is quadratic in $\phi$ and $\theta$ field and can therefore be solved exactly. These two field are canonical, which in particular means the following commutator
\be
[\phi(x,y), \partial_{x'} \theta(x',y')] = i \delta(x-x') \delta(y-y').
\label{eq42}
\ee
We use it to write equation of motion for $\partial_t \phi(x,y)$, and then one for $\partial_t \partial_x \theta(x,y)$, and then Fourier transform to frequency and wave vector space to find the dispersion relation of SDW excitations
\be
\omega^2_k(p_x, \vec{p}_\perp) = v^2 p_x^2 + c_y^2 (\vec{p}_\perp)^2 + m_k^2 .
\label{eq43}
\ee
Notice that $k$ in $\omega_k$ is the index reminding of the $k$-th plateau. Inside the plateau the excitations are massive (gapped), with the mass $m_k$ given by \eqref{eq41}, while in the incommensuare SDW phase the mass is absent ($m_k = 0$) and the excitations are linearly-dispersing phasons with very different velocities $v$ ($c_y$) along the chain (perpendicular to the chain) directions, correspondingly.

Eq.\eqref{eq43} is expected to work inside the ordered SDW (or plateau) phase at temperatures $T$ below $v \Lambda_{\rm sdw}$. That is, it represents the low-energy, acoustic, part of the phason dispersion. 


\subsection{Critical behavior of the ordering wave vector across C-IC transition}

Ref.\cite{extreme}, Section III.D.6, shows that the incommensurate correction $\delta Q = \pi \delta$ to the ordering wave vector $Q$ exhibits a singular dependence on the magnetic field. 
Specifically, deviation from the commensurability is described in terms of dilute gas of solitons which interpolate between the degenerate ground states of the classical sine-Gordon model \eqref{eq36}. In the low-density limit (that is, near the commensurate plateaux states) the solitons repel each other with potential that decays exponentially with the distance between them. This translates into the inverse log behavior of $\delta Q$,
\be
\delta Q = \frac{-2\pi}{w \ln[|h - h_{\nu,k}|/\Delta h]}
\ee
where $w$ is the width of the soliton. Notice that $\delta Q$ deviates from 0 with an infinite slope. This feature is clearly seen in Figure \ref{fig:plateaux}e, both near the C-IC transition from the AFM to LSDW phase and also near the C-IC transition from LSDW to the 1/3 magnetization plateau state. 

\section{S5: Magnetic field dependence of the AFM Bragg peak from neutron diffraction}
\begin{figure*}[!ht]
     \includegraphics[width=0.5\textwidth]{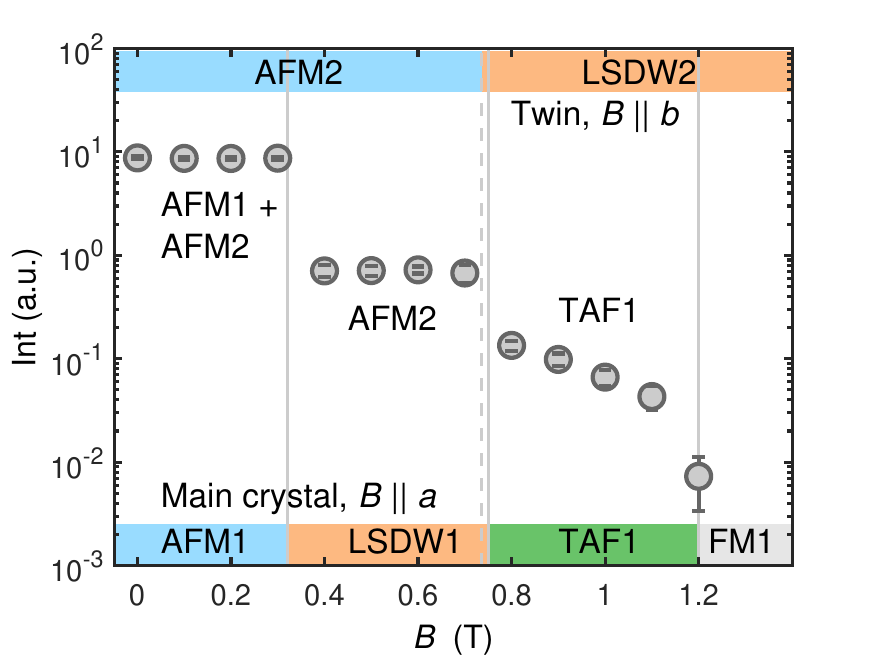}
   \caption{~Intensity of $(0,0,1)$ Bragg peak measured as a function of magnetic field using neutron diffraction~\cite{nikitin_multiple_2021}. Dots show the measured intensity, color-shaded areas below and above represent different phases for field applied along $a$ and $b$ axis correspondingly (see details in the main text). Grey solid (dotted) lines demonstrate the critical fields for the field applied along $a$ ($b$) axis.
    }
    \label{figsm:001_vs_field}
\end{figure*}
In this section we briefly discuss the magnetic field dependence of $(0,0,1)$ Bragg peak, which was measured by neutron diffraction (see Sec.~S1 of SM in Ref.~\cite{nikitin_multiple_2021} for the details) and which is important in the context of new phase transition at $B~\approx~1$~T.
AFM spin chains in \ybalo\ are aligned along the $c$ axis resulting in a Bragg peak at $\mathbf{Q} = (0,0,1)$ at zero field~\cite{RADHAKRISHNA1981, wu_antiferromagnetic_2019}. Moreover, the TAF phase is also expected to yield magnetic Bragg peak at the same $\mathbf{Q}$ position~\cite{fan_quantumcricality_2020}.
The field dependence of this peak was measured in Ref.~\cite{nikitin_multiple_2021}, but the sample contained two twins, which have common $c$ axis, but permuted $a$ and $b$ direction. Magnetic field was applied along the $a$ axis of the main twin, which constituted $\approx 95$~\%\ of total mass of the sample. For the second twin, the field was aligned along the $b$ axis.

Due to the g-factor anisotropy, the easy axis of Yb moments lies within the $ab$ plane with 23.5$^{\circ}$ to the $a$ direction~\cite{wu_antiferromagnetic_2019}. Thus, application of magnetic field along $a$ and $b$ produces qualitatively similar response, because it is caused by projection of the magnetic field on the easy magnetization axis.

Field-induced evolution of magnetic order in \ybalo\ for $B\|a$ (of the main twin) along with intensity of $(0,0,1)$ peak are summarized in Fig.~\ref{figsm:001_vs_field}. Intensity of $(0,0,1)$ is strong and constant below the first critical field, $B_{\rm c} = 0.32$~T. Above this field, it drops rapidly and remains constant up to $\approx\ 0.75$~T, which is close to the LSDW-TAF phase boundary for $B\|a$. Note that the field region between 0.32 and 0.75~T corresponds to the LSDW phase of the main twin, and no intensity is expected at $(0,0,1)$. However, this field region coincides with the AFM phase of the second twin, which has the first critical field of $B_{\rm c}/\mathrm{tan}(23.5^{\circ}) = 0.74$~T. Thus, we interpret the presence of the finite intensity between 0.32 and 0.75~T as signal from AFM Bragg peak of the second twin. At fields above 0.75~T the second twin is within the LSDW phase, and does not contribute to the intensity at $(0,0,1)$ and thus we can interpret the signal at $B > 0.74$~T (first critical field of the second twin) as the signal from the main twin. It shows a continuous decrease and disappear below the detection limit at $\approx~1.2$~T. 
This result suggest presence of a long-range magnetic order between 0.75~T and saturation field. Such behavior is consistent with formation of the TAF phase, as was proposed by theory~\cite{fan_quantumcricality_2020, fan_role_2020}.


\begin{thebibliography}{41}%
\makeatletter
\providecommand \@ifxundefined [1]{%
 \@ifx{#1\undefined}
}%
\providecommand \@ifnum [1]{%
 \ifnum #1\expandafter \@firstoftwo
 \else \expandafter \@secondoftwo
 \fi
}%
\providecommand \@ifx [1]{%
 \ifx #1\expandafter \@firstoftwo
 \else \expandafter \@secondoftwo
 \fi
}%
\providecommand \natexlab [1]{#1}%
\providecommand \enquote  [1]{``#1''}%
\providecommand \bibnamefont  [1]{#1}%
\providecommand \bibfnamefont [1]{#1}%
\providecommand \citenamefont [1]{#1}%
\providecommand \href@noop [0]{\@secondoftwo}%
\providecommand \href [0]{\begingroup \@sanitize@url \@href}%
\providecommand \@href[1]{\@@startlink{#1}\@@href}%
\providecommand \@@href[1]{\endgroup#1\@@endlink}%
\providecommand \@sanitize@url [0]{\catcode `\\12\catcode `\$12\catcode
  `\&12\catcode `\#12\catcode `\^12\catcode `\_12\catcode `\%12\relax}%
\providecommand \@@startlink[1]{}%
\providecommand \@@endlink[0]{}%
\providecommand \url  [0]{\begingroup\@sanitize@url \@url }%
\providecommand \@url [1]{\endgroup\@href {#1}{\urlprefix }}%
\providecommand \urlprefix  [0]{URL }%
\providecommand \Eprint [0]{\href }%
\providecommand \doibase [0]{https://doi.org/}%
\providecommand \selectlanguage [0]{\@gobble}%
\providecommand \bibinfo  [0]{\@secondoftwo}%
\providecommand \bibfield  [0]{\@secondoftwo}%
\providecommand \translation [1]{[#1]}%
\providecommand \BibitemOpen [0]{}%
\providecommand \bibitemStop [0]{}%
\providecommand \bibitemNoStop [0]{.\EOS\space}%
\providecommand \EOS [0]{\spacefactor3000\relax}%
\providecommand \BibitemShut  [1]{\csname bibitem#1\endcsname}%
\let\auto@bib@innerbib\@empty
\bibitem [{\citenamefont {Savary}\ and\ \citenamefont
  {Balents}(2016)}]{Savary2017}%
  \BibitemOpen
  \bibfield  {author} {\bibinfo {author} {\bibfnamefont {L.}~\bibnamefont
  {Savary}}\ and\ \bibinfo {author} {\bibfnamefont {L.}~\bibnamefont
  {Balents}},\ }\href {https://doi.org/10.1088/0034-4885/80/1/016502}
  {\bibfield  {journal} {\bibinfo  {journal} {Reports on Progress in Physics}\
  }\textbf {\bibinfo {volume} {80}},\ \bibinfo {pages} {016502} (\bibinfo
  {year} {2016})}\BibitemShut {NoStop}%
\bibitem [{\citenamefont {Starykh}(2015)}]{Starykh2015}%
  \BibitemOpen
  \bibfield  {author} {\bibinfo {author} {\bibfnamefont {O.~A.}\ \bibnamefont
  {Starykh}},\ }\bibfield  {journal} {\bibinfo  {journal} {Reports on Progress
  in Physics}\ }\textbf {\bibinfo {volume} {78}},\ \href
  {https://doi.org/10.1088/0034-4885/78/5/052502}
  {10.1088/0034-4885/78/5/052502} (\bibinfo {year} {2015})\BibitemShut
  {NoStop}%
\bibitem [{\citenamefont {Giamarchi}(2003)}]{Giamarchi_book}%
  \BibitemOpen
  \bibfield  {author} {\bibinfo {author} {\bibfnamefont {T.}~\bibnamefont
  {Giamarchi}},\ }\href {http://www.loc.gov/catdir/toc/fy051/2004299020.html}
  {\emph {\bibinfo {title} {Quantum physics in one dimension /}}},\ The
  international series of monographs on physics ;\ (\bibinfo  {publisher}
  {Clarendon ;},\ \bibinfo {address} {Oxford :},\ \bibinfo {year}
  {2003.})\BibitemShut {NoStop}%
\bibitem [{\citenamefont {Zamolodchikov}(1989)}]{zamolodchikov1989}%
  \BibitemOpen
  \bibfield  {author} {\bibinfo {author} {\bibfnamefont {A.~B.}\ \bibnamefont
  {Zamolodchikov}},\ }\href {https://doi.org/10.1142/S0217751X8900176X}
  {\bibfield  {journal} {\bibinfo  {journal} {International Journal of Modern
  Physics A}\ }\textbf {\bibinfo {volume} {04}},\ \bibinfo {pages} {4235}
  (\bibinfo {year} {1989})},\ \Eprint
  {https://arxiv.org/abs/https://doi.org/10.1142/S0217751X8900176X}
  {https://doi.org/10.1142/S0217751X8900176X} \BibitemShut {NoStop}%
\bibitem [{\citenamefont {Coldea}\ \emph {et~al.}(2010)\citenamefont {Coldea},
  \citenamefont {Tennant}, \citenamefont {Wheeler}, \citenamefont {Wawrzynska},
  \citenamefont {Prabhakaran}, \citenamefont {Telling}, \citenamefont
  {Habicht}, \citenamefont {Smeibidl},\ and\ \citenamefont
  {Kiefer}}]{coldea2010}%
  \BibitemOpen
  \bibfield  {author} {\bibinfo {author} {\bibfnamefont {R.}~\bibnamefont
  {Coldea}}, \bibinfo {author} {\bibfnamefont {D.~A.}\ \bibnamefont {Tennant}},
  \bibinfo {author} {\bibfnamefont {E.~M.}\ \bibnamefont {Wheeler}}, \bibinfo
  {author} {\bibfnamefont {E.}~\bibnamefont {Wawrzynska}}, \bibinfo {author}
  {\bibfnamefont {D.}~\bibnamefont {Prabhakaran}}, \bibinfo {author}
  {\bibfnamefont {M.}~\bibnamefont {Telling}}, \bibinfo {author} {\bibfnamefont
  {K.}~\bibnamefont {Habicht}}, \bibinfo {author} {\bibfnamefont
  {P.}~\bibnamefont {Smeibidl}},\ and\ \bibinfo {author} {\bibfnamefont
  {K.}~\bibnamefont {Kiefer}},\ }\href
  {https://doi.org/10.1126/science.1180085} {\bibfield  {journal} {\bibinfo
  {journal} {Science}\ }\textbf {\bibinfo {volume} {327}},\ \bibinfo {pages}
  {177} (\bibinfo {year} {2010})},\ \Eprint
  {https://arxiv.org/abs/https://www.science.org/doi/pdf/10.1126/science.1180085}
  {https://www.science.org/doi/pdf/10.1126/science.1180085} \BibitemShut
  {NoStop}%
\bibitem [{\citenamefont {Wu}\ \emph {et~al.}(2014)\citenamefont {Wu},
  \citenamefont {Kormos},\ and\ \citenamefont {Si}}]{Wu2014}%
  \BibitemOpen
  \bibfield  {author} {\bibinfo {author} {\bibfnamefont {J.}~\bibnamefont
  {Wu}}, \bibinfo {author} {\bibfnamefont {M.}~\bibnamefont {Kormos}},\ and\
  \bibinfo {author} {\bibfnamefont {Q.}~\bibnamefont {Si}},\ }\href
  {https://doi.org/10.1103/PhysRevLett.113.247201} {\bibfield  {journal}
  {\bibinfo  {journal} {Phys. Rev. Lett.}\ }\textbf {\bibinfo {volume} {113}},\
  \bibinfo {pages} {247201} (\bibinfo {year} {2014})}\BibitemShut {NoStop}%
\bibitem [{\citenamefont {Wang}\ \emph {et~al.}(2018)\citenamefont {Wang},
  \citenamefont {Wu}, \citenamefont {Yang}, \citenamefont {Bera}, \citenamefont
  {Kamenskyi}, \citenamefont {Islam}, \citenamefont {Xu}, \citenamefont {Law},
  \citenamefont {Lake}, \citenamefont {Wu},\ and\ \citenamefont
  {Loidl}}]{Wang2018}%
  \BibitemOpen
  \bibfield  {author} {\bibinfo {author} {\bibfnamefont {Z.}~\bibnamefont
  {Wang}}, \bibinfo {author} {\bibfnamefont {J.}~\bibnamefont {Wu}}, \bibinfo
  {author} {\bibfnamefont {W.}~\bibnamefont {Yang}}, \bibinfo {author}
  {\bibfnamefont {A.~K.}\ \bibnamefont {Bera}}, \bibinfo {author}
  {\bibfnamefont {D.}~\bibnamefont {Kamenskyi}}, \bibinfo {author}
  {\bibfnamefont {A.~T. M.~N.}\ \bibnamefont {Islam}}, \bibinfo {author}
  {\bibfnamefont {S.}~\bibnamefont {Xu}}, \bibinfo {author} {\bibfnamefont
  {J.~M.}\ \bibnamefont {Law}}, \bibinfo {author} {\bibfnamefont
  {B.}~\bibnamefont {Lake}}, \bibinfo {author} {\bibfnamefont {C.}~\bibnamefont
  {Wu}},\ and\ \bibinfo {author} {\bibfnamefont {A.}~\bibnamefont {Loidl}},\
  }\href {https://doi.org/10.1038/nature25466} {\bibfield  {journal} {\bibinfo
  {journal} {Nature}\ }\textbf {\bibinfo {volume} {554}},\ \bibinfo {pages}
  {219} (\bibinfo {year} {2018})}\BibitemShut {NoStop}%
\bibitem [{\citenamefont {Wang}\ \emph {et~al.}(2019)\citenamefont {Wang},
  \citenamefont {Schmidt}, \citenamefont {Loidl}, \citenamefont {Wu},
  \citenamefont {Zou}, \citenamefont {Yang}, \citenamefont {Dong},
  \citenamefont {Kohama}, \citenamefont {Kindo}, \citenamefont {Gorbunov},
  \citenamefont {Niesen}, \citenamefont {Breunig}, \citenamefont {Engelmayer},\
  and\ \citenamefont {Lorenz}}]{Wang2019}%
  \BibitemOpen
  \bibfield  {author} {\bibinfo {author} {\bibfnamefont {Z.}~\bibnamefont
  {Wang}}, \bibinfo {author} {\bibfnamefont {M.}~\bibnamefont {Schmidt}},
  \bibinfo {author} {\bibfnamefont {A.}~\bibnamefont {Loidl}}, \bibinfo
  {author} {\bibfnamefont {J.}~\bibnamefont {Wu}}, \bibinfo {author}
  {\bibfnamefont {H.}~\bibnamefont {Zou}}, \bibinfo {author} {\bibfnamefont
  {W.}~\bibnamefont {Yang}}, \bibinfo {author} {\bibfnamefont {C.}~\bibnamefont
  {Dong}}, \bibinfo {author} {\bibfnamefont {Y.}~\bibnamefont {Kohama}},
  \bibinfo {author} {\bibfnamefont {K.}~\bibnamefont {Kindo}}, \bibinfo
  {author} {\bibfnamefont {D.~I.}\ \bibnamefont {Gorbunov}}, \bibinfo {author}
  {\bibfnamefont {S.}~\bibnamefont {Niesen}}, \bibinfo {author} {\bibfnamefont
  {O.}~\bibnamefont {Breunig}}, \bibinfo {author} {\bibfnamefont
  {J.}~\bibnamefont {Engelmayer}},\ and\ \bibinfo {author} {\bibfnamefont
  {T.}~\bibnamefont {Lorenz}},\ }\href
  {https://doi.org/10.1103/PhysRevLett.123.067202} {\bibfield  {journal}
  {\bibinfo  {journal} {Phys. Rev. Lett.}\ }\textbf {\bibinfo {volume} {123}},\
  \bibinfo {pages} {067202} (\bibinfo {year} {2019})}\BibitemShut {NoStop}%
\bibitem [{\citenamefont {Yang}\ \emph {et~al.}(2023)\citenamefont {Yang},
  \citenamefont {Xie}, \citenamefont {Nikitin}, \citenamefont {Wu},\ and\
  \citenamefont {Podlesnyak}}]{Yang2023}%
  \BibitemOpen
  \bibfield  {author} {\bibinfo {author} {\bibfnamefont {J.}~\bibnamefont
  {Yang}}, \bibinfo {author} {\bibfnamefont {T.}~\bibnamefont {Xie}}, \bibinfo
  {author} {\bibfnamefont {S.~E.}\ \bibnamefont {Nikitin}}, \bibinfo {author}
  {\bibfnamefont {J.}~\bibnamefont {Wu}},\ and\ \bibinfo {author}
  {\bibfnamefont {A.}~\bibnamefont {Podlesnyak}},\ }\href
  {https://doi.org/10.1103/PhysRevB.108.L020402} {\bibfield  {journal}
  {\bibinfo  {journal} {Phys. Rev. B}\ }\textbf {\bibinfo {volume} {108}},\
  \bibinfo {pages} {L020402} (\bibinfo {year} {2023})}\BibitemShut {NoStop}%
\bibitem [{\citenamefont {Halati}\ \emph {et~al.}(2023)\citenamefont {Halati},
  \citenamefont {Wang}, \citenamefont {Lorenz}, \citenamefont {Kollath},\ and\
  \citenamefont {Bernier}}]{Halati2023}%
  \BibitemOpen
  \bibfield  {author} {\bibinfo {author} {\bibfnamefont {C.-M.}\ \bibnamefont
  {Halati}}, \bibinfo {author} {\bibfnamefont {Z.}~\bibnamefont {Wang}},
  \bibinfo {author} {\bibfnamefont {T.}~\bibnamefont {Lorenz}}, \bibinfo
  {author} {\bibfnamefont {C.}~\bibnamefont {Kollath}},\ and\ \bibinfo {author}
  {\bibfnamefont {J.-S.}\ \bibnamefont {Bernier}},\ }\bibfield  {journal}
  {\bibinfo  {journal} {Phys. Rev. B}\ }\textbf {\bibinfo {volume} {108}},\
  \href {https://doi.org/10.1103/PhysRevB.108.224429}
  {10.1103/PhysRevB.108.224429} (\bibinfo {year} {2023})\BibitemShut {NoStop}%
\bibitem [{\citenamefont {Wang}\ \emph {et~al.}(2024)\citenamefont {Wang},
  \citenamefont {Halati}, \citenamefont {Bernier}, \citenamefont {Ponomaryov},
  \citenamefont {Gorbunov}, \citenamefont {Niesen}, \citenamefont {Breunig},
  \citenamefont {Klopf}, \citenamefont {Zvyagin}, \citenamefont {Lorenz},
  \citenamefont {Loidl},\ and\ \citenamefont {Kollath}}]{Wang2024}%
  \BibitemOpen
  \bibfield  {author} {\bibinfo {author} {\bibfnamefont {Z.}~\bibnamefont
  {Wang}}, \bibinfo {author} {\bibfnamefont {C.-M.}\ \bibnamefont {Halati}},
  \bibinfo {author} {\bibfnamefont {J.-S.}\ \bibnamefont {Bernier}}, \bibinfo
  {author} {\bibfnamefont {A.}~\bibnamefont {Ponomaryov}}, \bibinfo {author}
  {\bibfnamefont {D.~I.}\ \bibnamefont {Gorbunov}}, \bibinfo {author}
  {\bibfnamefont {S.}~\bibnamefont {Niesen}}, \bibinfo {author} {\bibfnamefont
  {O.}~\bibnamefont {Breunig}}, \bibinfo {author} {\bibfnamefont {J.~M.}\
  \bibnamefont {Klopf}}, \bibinfo {author} {\bibfnamefont {S.}~\bibnamefont
  {Zvyagin}}, \bibinfo {author} {\bibfnamefont {T.}~\bibnamefont {Lorenz}},
  \bibinfo {author} {\bibfnamefont {A.}~\bibnamefont {Loidl}},\ and\ \bibinfo
  {author} {\bibfnamefont {C.}~\bibnamefont {Kollath}},\ }\bibfield  {journal}
  {\bibinfo  {journal} {Nature}\ }\textbf {\bibinfo {volume} {631}},\ \href
  {https://doi.org/10.1038/s41586-024-07599-3} {10.1038/s41586-024-07599-3}
  (\bibinfo {year} {2024})\BibitemShut {NoStop}%
\bibitem [{\citenamefont {Klanj\ifmmode~\check{s}\else \v{s}\fi{}ek}\ \emph
  {et~al.}(2015)\citenamefont {Klanj\ifmmode~\check{s}\else \v{s}\fi{}ek},
  \citenamefont {Horvati\ifmmode~\acute{c}\else \'{c}\fi{}}, \citenamefont
  {Kr\"amer}, \citenamefont {Mukhopadhyay}, \citenamefont {Mayaffre},
  \citenamefont {Berthier}, \citenamefont {Can\'evet}, \citenamefont {Grenier},
  \citenamefont {Lejay},\ and\ \citenamefont {Orignac}}]{Klanjsek2015}%
  \BibitemOpen
  \bibfield  {author} {\bibinfo {author} {\bibfnamefont {M.}~\bibnamefont
  {Klanj\ifmmode~\check{s}\else \v{s}\fi{}ek}}, \bibinfo {author}
  {\bibfnamefont {M.}~\bibnamefont {Horvati\ifmmode~\acute{c}\else
  \'{c}\fi{}}}, \bibinfo {author} {\bibfnamefont {S.}~\bibnamefont {Kr\"amer}},
  \bibinfo {author} {\bibfnamefont {S.}~\bibnamefont {Mukhopadhyay}}, \bibinfo
  {author} {\bibfnamefont {H.}~\bibnamefont {Mayaffre}}, \bibinfo {author}
  {\bibfnamefont {C.}~\bibnamefont {Berthier}}, \bibinfo {author}
  {\bibfnamefont {E.}~\bibnamefont {Can\'evet}}, \bibinfo {author}
  {\bibfnamefont {B.}~\bibnamefont {Grenier}}, \bibinfo {author} {\bibfnamefont
  {P.}~\bibnamefont {Lejay}},\ and\ \bibinfo {author} {\bibfnamefont
  {E.}~\bibnamefont {Orignac}},\ }\href
  {https://doi.org/10.1103/PhysRevB.92.060408} {\bibfield  {journal} {\bibinfo
  {journal} {Phys. Rev. B}\ }\textbf {\bibinfo {volume} {92}},\ \bibinfo
  {pages} {060408} (\bibinfo {year} {2015})}\BibitemShut {NoStop}%
\bibitem [{\citenamefont {Takayoshi}\ \emph {et~al.}(2023)\citenamefont
  {Takayoshi}, \citenamefont {Faure}, \citenamefont {Simonet}, \citenamefont
  {Grenier}, \citenamefont {Petit}, \citenamefont {Ollivier}, \citenamefont
  {Lejay},\ and\ \citenamefont {Giamarchi}}]{Takayoshi2023}%
  \BibitemOpen
  \bibfield  {author} {\bibinfo {author} {\bibfnamefont {S.}~\bibnamefont
  {Takayoshi}}, \bibinfo {author} {\bibfnamefont {Q.}~\bibnamefont {Faure}},
  \bibinfo {author} {\bibfnamefont {V.}~\bibnamefont {Simonet}}, \bibinfo
  {author} {\bibfnamefont {B.}~\bibnamefont {Grenier}}, \bibinfo {author}
  {\bibfnamefont {S.}~\bibnamefont {Petit}}, \bibinfo {author} {\bibfnamefont
  {J.}~\bibnamefont {Ollivier}}, \bibinfo {author} {\bibfnamefont
  {P.}~\bibnamefont {Lejay}},\ and\ \bibinfo {author} {\bibfnamefont
  {T.}~\bibnamefont {Giamarchi}},\ }\bibfield  {journal} {\bibinfo  {journal}
  {Phys. Rev. Research}\ }\textbf {\bibinfo {volume} {5}},\ \href
  {https://doi.org/10.1103/PhysRevResearch.5.023205}
  {10.1103/PhysRevResearch.5.023205} (\bibinfo {year} {2023})\BibitemShut
  {NoStop}%
\bibitem [{\citenamefont {Wu}\ \emph {et~al.}(2019{\natexlab{a}})\citenamefont
  {Wu}, \citenamefont {Nikitin}, \citenamefont {Brando}, \citenamefont
  {Vasylechko}, \citenamefont {Ehlers}, \citenamefont {Frontzek}, \citenamefont
  {Savici}, \citenamefont {Sala}, \citenamefont {Christianson}, \citenamefont
  {Lumsden},\ and\ \citenamefont {Podlesnyak}}]{wu_antiferromagnetic_2019}%
  \BibitemOpen
  \bibfield  {author} {\bibinfo {author} {\bibfnamefont {L.~S.}\ \bibnamefont
  {Wu}}, \bibinfo {author} {\bibfnamefont {S.~E.}\ \bibnamefont {Nikitin}},
  \bibinfo {author} {\bibfnamefont {M.}~\bibnamefont {Brando}}, \bibinfo
  {author} {\bibfnamefont {L.}~\bibnamefont {Vasylechko}}, \bibinfo {author}
  {\bibfnamefont {G.}~\bibnamefont {Ehlers}}, \bibinfo {author} {\bibfnamefont
  {M.}~\bibnamefont {Frontzek}}, \bibinfo {author} {\bibfnamefont {A.~T.}\
  \bibnamefont {Savici}}, \bibinfo {author} {\bibfnamefont {G.}~\bibnamefont
  {Sala}}, \bibinfo {author} {\bibfnamefont {A.~D.}\ \bibnamefont
  {Christianson}}, \bibinfo {author} {\bibfnamefont {M.~D.}\ \bibnamefont
  {Lumsden}},\ and\ \bibinfo {author} {\bibfnamefont {A.}~\bibnamefont
  {Podlesnyak}},\ }\href {https://doi.org/10.1103/PhysRevB.99.195117}
  {\bibfield  {journal} {\bibinfo  {journal} {Phys. Rev. B}\ }\textbf {\bibinfo
  {volume} {99}},\ \bibinfo {pages} {195117} (\bibinfo {year}
  {2019}{\natexlab{a}})}\BibitemShut {NoStop}%
\bibitem [{\citenamefont {Wu}\ \emph {et~al.}(2016)\citenamefont {Wu},
  \citenamefont {Gannon}, \citenamefont {Zaliznyak}, \citenamefont {Tsvelik},
  \citenamefont {Brockmann}, \citenamefont {Caux}, \citenamefont {Kim},
  \citenamefont {Qiu}, \citenamefont {Copley}, \citenamefont {Ehlers},
  \citenamefont {Podlesnyak},\ and\ \citenamefont
  {Aronson}}]{Wu-Zaliznyak2016}%
  \BibitemOpen
  \bibfield  {author} {\bibinfo {author} {\bibfnamefont {L.~S.}\ \bibnamefont
  {Wu}}, \bibinfo {author} {\bibfnamefont {W.~J.}\ \bibnamefont {Gannon}},
  \bibinfo {author} {\bibfnamefont {I.~A.}\ \bibnamefont {Zaliznyak}}, \bibinfo
  {author} {\bibfnamefont {A.~M.}\ \bibnamefont {Tsvelik}}, \bibinfo {author}
  {\bibfnamefont {M.}~\bibnamefont {Brockmann}}, \bibinfo {author}
  {\bibfnamefont {J.-S.}\ \bibnamefont {Caux}}, \bibinfo {author}
  {\bibfnamefont {M.~S.}\ \bibnamefont {Kim}}, \bibinfo {author} {\bibfnamefont
  {Y.}~\bibnamefont {Qiu}}, \bibinfo {author} {\bibfnamefont {J.~R.~D.}\
  \bibnamefont {Copley}}, \bibinfo {author} {\bibfnamefont {G.}~\bibnamefont
  {Ehlers}}, \bibinfo {author} {\bibfnamefont {A.}~\bibnamefont {Podlesnyak}},\
  and\ \bibinfo {author} {\bibfnamefont {M.~C.}\ \bibnamefont {Aronson}},\
  }\href {https://doi.org/10.1126/science.aaf0981} {\bibfield  {journal}
  {\bibinfo  {journal} {Science}\ }\textbf {\bibinfo {volume} {352}},\ \bibinfo
  {pages} {1206} (\bibinfo {year} {2016})},\ \Eprint
  {https://arxiv.org/abs/https://www.science.org/doi/pdf/10.1126/science.aaf0981}
  {https://www.science.org/doi/pdf/10.1126/science.aaf0981} \BibitemShut
  {NoStop}%
\bibitem [{\citenamefont {Wu}\ \emph {et~al.}(2019{\natexlab{b}})\citenamefont
  {Wu}, \citenamefont {Nikitin}, \citenamefont {Wang}, \citenamefont {Zhu},
  \citenamefont {Batista}, \citenamefont {Tsvelik}, \citenamefont {Samarakoon},
  \citenamefont {Tennant}, \citenamefont {Brando}, \citenamefont {Vasylechko},
  \citenamefont {Frontzek}, \citenamefont {Savici}, \citenamefont {Sala},
  \citenamefont {Ehlers}, \citenamefont {Christianson}, \citenamefont
  {Lumsden},\ and\ \citenamefont {Podlesnyak}}]{wu_tomonagaluttinger_2019}%
  \BibitemOpen
  \bibfield  {author} {\bibinfo {author} {\bibfnamefont {L.~S.}\ \bibnamefont
  {Wu}}, \bibinfo {author} {\bibfnamefont {S.~E.}\ \bibnamefont {Nikitin}},
  \bibinfo {author} {\bibfnamefont {Z.}~\bibnamefont {Wang}}, \bibinfo {author}
  {\bibfnamefont {W.}~\bibnamefont {Zhu}}, \bibinfo {author} {\bibfnamefont
  {C.~D.}\ \bibnamefont {Batista}}, \bibinfo {author} {\bibfnamefont {A.~M.}\
  \bibnamefont {Tsvelik}}, \bibinfo {author} {\bibfnamefont {A.~M.}\
  \bibnamefont {Samarakoon}}, \bibinfo {author} {\bibfnamefont {D.~A.}\
  \bibnamefont {Tennant}}, \bibinfo {author} {\bibfnamefont {M.}~\bibnamefont
  {Brando}}, \bibinfo {author} {\bibfnamefont {L.}~\bibnamefont {Vasylechko}},
  \bibinfo {author} {\bibfnamefont {M.}~\bibnamefont {Frontzek}}, \bibinfo
  {author} {\bibfnamefont {A.~T.}\ \bibnamefont {Savici}}, \bibinfo {author}
  {\bibfnamefont {G.}~\bibnamefont {Sala}}, \bibinfo {author} {\bibfnamefont
  {G.}~\bibnamefont {Ehlers}}, \bibinfo {author} {\bibfnamefont {A.~D.}\
  \bibnamefont {Christianson}}, \bibinfo {author} {\bibfnamefont {M.~D.}\
  \bibnamefont {Lumsden}},\ and\ \bibinfo {author} {\bibfnamefont
  {A.}~\bibnamefont {Podlesnyak}},\ }\href
  {https://doi.org/10.1038/s41467-019-08485-7} {\bibfield  {journal} {\bibinfo
  {journal} {Nature Communications}\ }\textbf {\bibinfo {volume} {10}},\
  \bibinfo {pages} {698} (\bibinfo {year} {2019}{\natexlab{b}})}\BibitemShut
  {NoStop}%
\bibitem [{\citenamefont {Nikitin}\ \emph {et~al.}(2021)\citenamefont
  {Nikitin}, \citenamefont {Nishimoto}, \citenamefont {Fan}, \citenamefont
  {Wu}, \citenamefont {Wu}, \citenamefont {Sukhanov}, \citenamefont {Brando},
  \citenamefont {Pavlovskii}, \citenamefont {Xu}, \citenamefont {Vasylechko},
  \citenamefont {Yu},\ and\ \citenamefont
  {Podlesnyak}}]{nikitin_multiple_2021}%
  \BibitemOpen
  \bibfield  {author} {\bibinfo {author} {\bibfnamefont {S.~E.}\ \bibnamefont
  {Nikitin}}, \bibinfo {author} {\bibfnamefont {S.}~\bibnamefont {Nishimoto}},
  \bibinfo {author} {\bibfnamefont {Y.}~\bibnamefont {Fan}}, \bibinfo {author}
  {\bibfnamefont {J.}~\bibnamefont {Wu}}, \bibinfo {author} {\bibfnamefont
  {L.~S.}\ \bibnamefont {Wu}}, \bibinfo {author} {\bibfnamefont {A.~S.}\
  \bibnamefont {Sukhanov}}, \bibinfo {author} {\bibfnamefont {M.}~\bibnamefont
  {Brando}}, \bibinfo {author} {\bibfnamefont {N.~S.}\ \bibnamefont
  {Pavlovskii}}, \bibinfo {author} {\bibfnamefont {J.}~\bibnamefont {Xu}},
  \bibinfo {author} {\bibfnamefont {L.}~\bibnamefont {Vasylechko}}, \bibinfo
  {author} {\bibfnamefont {R.}~\bibnamefont {Yu}},\ and\ \bibinfo {author}
  {\bibfnamefont {A.}~\bibnamefont {Podlesnyak}},\ }\href
  {https://doi.org/10.1038/s41467-021-23585-z} {\bibfield  {journal} {\bibinfo
  {journal} {Nature Communications}\ }\textbf {\bibinfo {volume} {12}},\
  \bibinfo {pages} {3599} (\bibinfo {year} {2021})}\BibitemShut {NoStop}%
\bibitem [{\citenamefont {Buryy}\ \emph {et~al.}(2010)\citenamefont {Buryy},
  \citenamefont {Zhydachevskii}, \citenamefont {Vasylechko}, \citenamefont
  {Sugak}, \citenamefont {Martynyuk}, \citenamefont {Ubizskii},\ and\
  \citenamefont {Becker}}]{Buryy_growth_2010}%
  \BibitemOpen
  \bibfield  {author} {\bibinfo {author} {\bibfnamefont {O.}~\bibnamefont
  {Buryy}}, \bibinfo {author} {\bibfnamefont {Y.}~\bibnamefont
  {Zhydachevskii}}, \bibinfo {author} {\bibfnamefont {L.}~\bibnamefont
  {Vasylechko}}, \bibinfo {author} {\bibfnamefont {D.}~\bibnamefont {Sugak}},
  \bibinfo {author} {\bibfnamefont {N.}~\bibnamefont {Martynyuk}}, \bibinfo
  {author} {\bibfnamefont {S.}~\bibnamefont {Ubizskii}},\ and\ \bibinfo
  {author} {\bibfnamefont {K.~D.}\ \bibnamefont {Becker}},\ }\href
  {https://doi.org/10.1088/0953-8984/22/5/055902} {\bibfield  {journal}
  {\bibinfo  {journal} {Journal of Physics: Condensed Matter}\ }\textbf
  {\bibinfo {volume} {22}},\ \bibinfo {pages} {055902} (\bibinfo {year}
  {2010})}\BibitemShut {NoStop}%
\bibitem [{\citenamefont {Ehlers}\ \emph {et~al.}(2022)\citenamefont {Ehlers},
  \citenamefont {Vasylechko}, \citenamefont {Medvecká},\ and\ \citenamefont
  {Sichelschmidt}}]{ehlers_esp_2022}%
  \BibitemOpen
  \bibfield  {author} {\bibinfo {author} {\bibfnamefont {D.}~\bibnamefont
  {Ehlers}}, \bibinfo {author} {\bibfnamefont {L.}~\bibnamefont {Vasylechko}},
  \bibinfo {author} {\bibfnamefont {Z.}~\bibnamefont {Medvecká}},\ and\
  \bibinfo {author} {\bibfnamefont {J.}~\bibnamefont {Sichelschmidt}},\ }\href
  {https://doi.org/10.1007/s00723-022-01483-x} {\bibfield  {journal} {\bibinfo
  {journal} {Appl Magn Reson.}\ }\textbf {\bibinfo {volume} {3}},\ \bibinfo
  {pages} {1399–1405} (\bibinfo {year} {2022})}\BibitemShut {NoStop}%
\bibitem [{\citenamefont {Radhakrishna}\ \emph {et~al.}(1981)\citenamefont
  {Radhakrishna}, \citenamefont {Hammann}, \citenamefont {Ocio}, \citenamefont
  {Pari},\ and\ \citenamefont {Allain}}]{RADHAKRISHNA1981}%
  \BibitemOpen
  \bibfield  {author} {\bibinfo {author} {\bibfnamefont {P.}~\bibnamefont
  {Radhakrishna}}, \bibinfo {author} {\bibfnamefont {J.}~\bibnamefont
  {Hammann}}, \bibinfo {author} {\bibfnamefont {M.}~\bibnamefont {Ocio}},
  \bibinfo {author} {\bibfnamefont {P.}~\bibnamefont {Pari}},\ and\ \bibinfo
  {author} {\bibfnamefont {Y.}~\bibnamefont {Allain}},\ }\href
  {https://doi.org/https://doi.org/10.1016/0038-1098(81)91181-9} {\bibfield
  {journal} {\bibinfo  {journal} {Solid State Communications}\ }\textbf
  {\bibinfo {volume} {37}},\ \bibinfo {pages} {813} (\bibinfo {year}
  {1981})}\BibitemShut {NoStop}%
\bibitem [{\citenamefont {Nikitin}\ \emph {et~al.}(2020)\citenamefont
  {Nikitin}, \citenamefont {Xie}, \citenamefont {Podlesnyak},\ and\
  \citenamefont {Zaliznyak}}]{nikitin_dimers_2020}%
  \BibitemOpen
  \bibfield  {author} {\bibinfo {author} {\bibfnamefont {S.~E.}\ \bibnamefont
  {Nikitin}}, \bibinfo {author} {\bibfnamefont {T.}~\bibnamefont {Xie}},
  \bibinfo {author} {\bibfnamefont {A.}~\bibnamefont {Podlesnyak}},\ and\
  \bibinfo {author} {\bibfnamefont {I.~A.}\ \bibnamefont {Zaliznyak}},\ }\href
  {https://doi.org/10.1103/PhysRevB.101.245150} {\bibfield  {journal} {\bibinfo
   {journal} {Phys. Rev. B}\ }\textbf {\bibinfo {volume} {101}},\ \bibinfo
  {pages} {245150} (\bibinfo {year} {2020})}\BibitemShut {NoStop}%
\bibitem [{\citenamefont {Fan}\ and\ \citenamefont {Yu}(2020)}]{fan_role_2020}%
  \BibitemOpen
  \bibfield  {author} {\bibinfo {author} {\bibfnamefont {Y.}~\bibnamefont
  {Fan}}\ and\ \bibinfo {author} {\bibfnamefont {R.}~\bibnamefont {Yu}},\
  }\href {https://doi.org/10.1088/1674-1056/ab820b} {\bibfield  {journal}
  {\bibinfo  {journal} {Chinese Physics B}\ }\textbf {\bibinfo {volume} {29}},\
  \bibinfo {pages} {057505} (\bibinfo {year} {2020})}\BibitemShut {NoStop}%
\bibitem [{\citenamefont {Okunishi}\ and\ \citenamefont
  {Suzuki}(2007)}]{Okunishi2007}%
  \BibitemOpen
  \bibfield  {author} {\bibinfo {author} {\bibfnamefont {K.}~\bibnamefont
  {Okunishi}}\ and\ \bibinfo {author} {\bibfnamefont {T.}~\bibnamefont
  {Suzuki}},\ }\href {https://doi.org/10.1103/PhysRevB.76.224411} {\bibfield
  {journal} {\bibinfo  {journal} {Phys. Rev. B}\ }\textbf {\bibinfo {volume}
  {76}},\ \bibinfo {pages} {224411} (\bibinfo {year} {2007})}\BibitemShut
  {NoStop}%
\bibitem [{\citenamefont {Fan}\ \emph {et~al.}(2020)\citenamefont {Fan},
  \citenamefont {Yang}, \citenamefont {Yu}, \citenamefont {Wu},\ and\
  \citenamefont {Yu}}]{fan_quantumcricality_2020}%
  \BibitemOpen
  \bibfield  {author} {\bibinfo {author} {\bibfnamefont {Y.}~\bibnamefont
  {Fan}}, \bibinfo {author} {\bibfnamefont {J.}~\bibnamefont {Yang}}, \bibinfo
  {author} {\bibfnamefont {W.}~\bibnamefont {Yu}}, \bibinfo {author}
  {\bibfnamefont {J.}~\bibnamefont {Wu}},\ and\ \bibinfo {author}
  {\bibfnamefont {R.}~\bibnamefont {Yu}},\ }\href
  {https://doi.org/10.1103/PhysRevResearch.2.013345} {\bibfield  {journal}
  {\bibinfo  {journal} {Phys. Rev. Research}\ }\textbf {\bibinfo {volume}
  {2}},\ \bibinfo {pages} {013345} (\bibinfo {year} {2020})}\BibitemShut
  {NoStop}%
\bibitem [{\citenamefont {Agrapidis}\ \emph {et~al.}(2019)\citenamefont
  {Agrapidis}, \citenamefont {van~den Brink},\ and\ \citenamefont
  {Nishimoto}}]{Agrapidis_incommensurate_2019}%
  \BibitemOpen
  \bibfield  {author} {\bibinfo {author} {\bibfnamefont {C.~E.}\ \bibnamefont
  {Agrapidis}}, \bibinfo {author} {\bibfnamefont {J.}~\bibnamefont {van~den
  Brink}},\ and\ \bibinfo {author} {\bibfnamefont {S.}~\bibnamefont
  {Nishimoto}},\ }\href {https://doi.org/10.1103/PhysRevB.99.224423} {\bibfield
   {journal} {\bibinfo  {journal} {Phys. Rev. B}\ }\textbf {\bibinfo {volume}
  {99}},\ \bibinfo {pages} {224423} (\bibinfo {year} {2019})}\BibitemShut
  {NoStop}%
\bibitem [{sm1()}]{sm1}%
  \BibitemOpen
  \href@noop {} {}\bibinfo {note} {See Supplemental Material at http://...
  which contains descriptions of the experimental methods and protocols, some
  further analysis and details about the theory. Supplemental Material also
  includes
  Refs.~\cite{noginov_role_2001,sakakibara_magnetometer_1994,kuechler_new_applications_2023,kuechler_smallest_2017,extreme,nematic,LB2008}.}\BibitemShut
  {Stop}%
\bibitem [{\citenamefont {Zapf}\ \emph {et~al.}(2008)\citenamefont {Zapf},
  \citenamefont {Correa}, \citenamefont {Sengupta}, \citenamefont {Batista},
  \citenamefont {Tsukamoto}, \citenamefont {Kawashima}, \citenamefont {Egan},
  \citenamefont {Pantea}, \citenamefont {Migliori}, \citenamefont {Betts},
  \citenamefont {Jaime},\ and\ \citenamefont {Paduan-Filho}}]{Zapf2008}%
  \BibitemOpen
  \bibfield  {author} {\bibinfo {author} {\bibfnamefont {V.~S.}\ \bibnamefont
  {Zapf}}, \bibinfo {author} {\bibfnamefont {V.~F.}\ \bibnamefont {Correa}},
  \bibinfo {author} {\bibfnamefont {P.}~\bibnamefont {Sengupta}}, \bibinfo
  {author} {\bibfnamefont {C.~D.}\ \bibnamefont {Batista}}, \bibinfo {author}
  {\bibfnamefont {M.}~\bibnamefont {Tsukamoto}}, \bibinfo {author}
  {\bibfnamefont {N.}~\bibnamefont {Kawashima}}, \bibinfo {author}
  {\bibfnamefont {P.}~\bibnamefont {Egan}}, \bibinfo {author} {\bibfnamefont
  {C.}~\bibnamefont {Pantea}}, \bibinfo {author} {\bibfnamefont
  {A.}~\bibnamefont {Migliori}}, \bibinfo {author} {\bibfnamefont {J.~B.}\
  \bibnamefont {Betts}}, \bibinfo {author} {\bibfnamefont {M.}~\bibnamefont
  {Jaime}},\ and\ \bibinfo {author} {\bibfnamefont {A.}~\bibnamefont
  {Paduan-Filho}},\ }\href {https://doi.org/10.1103/PhysRevB.77.020404}
  {\bibfield  {journal} {\bibinfo  {journal} {Phys. Rev. B}\ }\textbf {\bibinfo
  {volume} {77}},\ \bibinfo {pages} {020404} (\bibinfo {year}
  {2008})}\BibitemShut {NoStop}%
\bibitem [{\citenamefont {Miyata}\ \emph {et~al.}(2021)\citenamefont {Miyata},
  \citenamefont {Hikihara}, \citenamefont {Furukawa}, \citenamefont {Kremer},
  \citenamefont {Zherlitsyn},\ and\ \citenamefont {Wosnitza}}]{Miyata2021}%
  \BibitemOpen
  \bibfield  {author} {\bibinfo {author} {\bibfnamefont {A.}~\bibnamefont
  {Miyata}}, \bibinfo {author} {\bibfnamefont {T.}~\bibnamefont {Hikihara}},
  \bibinfo {author} {\bibfnamefont {S.}~\bibnamefont {Furukawa}}, \bibinfo
  {author} {\bibfnamefont {R.~K.}\ \bibnamefont {Kremer}}, \bibinfo {author}
  {\bibfnamefont {S.}~\bibnamefont {Zherlitsyn}},\ and\ \bibinfo {author}
  {\bibfnamefont {J.}~\bibnamefont {Wosnitza}},\ }\href
  {https://doi.org/10.1103/PhysRevB.103.014411} {\bibfield  {journal} {\bibinfo
   {journal} {Phys. Rev. B}\ }\textbf {\bibinfo {volume} {103}},\ \bibinfo
  {pages} {014411} (\bibinfo {year} {2021})}\BibitemShut {NoStop}%
\bibitem [{\citenamefont {Hess}(2019)}]{Hess2019}%
  \BibitemOpen
  \bibfield  {author} {\bibinfo {author} {\bibfnamefont {C.}~\bibnamefont
  {Hess}},\ }\href
  {https://doi.org/https://doi.org/10.1016/j.physrep.2019.02.004} {\bibfield
  {journal} {\bibinfo  {journal} {Physics Reports}\ }\textbf {\bibinfo {volume}
  {811}},\ \bibinfo {pages} {1} (\bibinfo {year} {2019})},\ \bibinfo {note}
  {heat transport of cuprate-based low-dimensional quantum magnets with strong
  exchange coupling}\BibitemShut {NoStop}%
\bibitem [{\citenamefont {Mokhtari}\ \emph {et~al.}(2025)\citenamefont
  {Mokhtari}, \citenamefont {Stockert}, \citenamefont {Nikitin}, \citenamefont
  {Vasylechko}, \citenamefont {Brando},\ and\ \citenamefont
  {Hassinger}}]{Mokhtari2025a}%
  \BibitemOpen
  \bibfield  {author} {\bibinfo {author} {\bibfnamefont {P.}~\bibnamefont
  {Mokhtari}}, \bibinfo {author} {\bibfnamefont {U.}~\bibnamefont {Stockert}},
  \bibinfo {author} {\bibfnamefont {S.~E.}\ \bibnamefont {Nikitin}}, \bibinfo
  {author} {\bibfnamefont {L.}~\bibnamefont {Vasylechko}}, \bibinfo {author}
  {\bibfnamefont {M.}~\bibnamefont {Brando}},\ and\ \bibinfo {author}
  {\bibfnamefont {E.}~\bibnamefont {Hassinger}},\ }\bibfield  {journal}
  {\bibinfo  {journal} {Journal of Applied Physics}\ }\href@noop {} {}
  (\bibinfo {year} {2025})\BibitemShut {NoStop}%
\bibitem [{Mok()}]{Mokhtari2025}%
  \BibitemOpen
  \href@noop {} {}\bibinfo {note} {P. Mokhtari et al., to be
  published.}\BibitemShut {Stop}%
\bibitem [{\citenamefont {Ranjith}\ \emph {et~al.}(2019)\citenamefont
  {Ranjith}, \citenamefont {Luther}, \citenamefont {Reimann}, \citenamefont
  {Schmidt}, \citenamefont {Schlender}, \citenamefont {Sichelschmidt},
  \citenamefont {Yasuoka}, \citenamefont {Strydom}, \citenamefont {Skourski},
  \citenamefont {Wosnitza}, \citenamefont {K\"uhne}, \citenamefont {Doert},\
  and\ \citenamefont {Baenitz}}]{RanjithPRB2019}%
  \BibitemOpen
  \bibfield  {author} {\bibinfo {author} {\bibfnamefont {K.~M.}\ \bibnamefont
  {Ranjith}}, \bibinfo {author} {\bibfnamefont {S.}~\bibnamefont {Luther}},
  \bibinfo {author} {\bibfnamefont {T.}~\bibnamefont {Reimann}}, \bibinfo
  {author} {\bibfnamefont {B.}~\bibnamefont {Schmidt}}, \bibinfo {author}
  {\bibfnamefont {P.}~\bibnamefont {Schlender}}, \bibinfo {author}
  {\bibfnamefont {J.}~\bibnamefont {Sichelschmidt}}, \bibinfo {author}
  {\bibfnamefont {H.}~\bibnamefont {Yasuoka}}, \bibinfo {author} {\bibfnamefont
  {A.~M.}\ \bibnamefont {Strydom}}, \bibinfo {author} {\bibfnamefont
  {Y.}~\bibnamefont {Skourski}}, \bibinfo {author} {\bibfnamefont
  {J.}~\bibnamefont {Wosnitza}}, \bibinfo {author} {\bibfnamefont
  {H.}~\bibnamefont {K\"uhne}}, \bibinfo {author} {\bibfnamefont
  {T.}~\bibnamefont {Doert}},\ and\ \bibinfo {author} {\bibfnamefont
  {M.}~\bibnamefont {Baenitz}},\ }\href
  {https://doi.org/10.1103/PhysRevB.100.224417} {\bibfield  {journal} {\bibinfo
   {journal} {Phys. Rev. B}\ }\textbf {\bibinfo {volume} {100}},\ \bibinfo
  {pages} {224417} (\bibinfo {year} {2019})}\BibitemShut {NoStop}%
\bibitem [{\citenamefont {Fortune}\ \emph {et~al.}(2009)\citenamefont
  {Fortune}, \citenamefont {Hannahs}, \citenamefont {Yoshida}, \citenamefont
  {Sherline}, \citenamefont {Ono}, \citenamefont {Tanaka},\ and\ \citenamefont
  {Takano}}]{fortune2009}%
  \BibitemOpen
  \bibfield  {author} {\bibinfo {author} {\bibfnamefont {N.~A.}\ \bibnamefont
  {Fortune}}, \bibinfo {author} {\bibfnamefont {S.~T.}\ \bibnamefont
  {Hannahs}}, \bibinfo {author} {\bibfnamefont {Y.}~\bibnamefont {Yoshida}},
  \bibinfo {author} {\bibfnamefont {T.~E.}\ \bibnamefont {Sherline}}, \bibinfo
  {author} {\bibfnamefont {T.}~\bibnamefont {Ono}}, \bibinfo {author}
  {\bibfnamefont {H.}~\bibnamefont {Tanaka}},\ and\ \bibinfo {author}
  {\bibfnamefont {Y.}~\bibnamefont {Takano}},\ }\href
  {https://doi.org/10.1103/PhysRevLett.102.257201} {\bibfield  {journal}
  {\bibinfo  {journal} {Phys. Rev. Lett.}\ }\textbf {\bibinfo {volume} {102}},\
  \bibinfo {pages} {257201} (\bibinfo {year} {2009})}\BibitemShut {NoStop}%
\bibitem [{\citenamefont {Facheris}\ \emph {et~al.}(2022)\citenamefont
  {Facheris}, \citenamefont {Povarov}, \citenamefont {Nabi}, \citenamefont
  {Mazzone}, \citenamefont {Lass}, \citenamefont {Roessli}, \citenamefont
  {Ressouche}, \citenamefont {Yan}, \citenamefont {Gvasaliya},\ and\
  \citenamefont {Zheludev}}]{Facheris2022}%
  \BibitemOpen
  \bibfield  {author} {\bibinfo {author} {\bibfnamefont {L.}~\bibnamefont
  {Facheris}}, \bibinfo {author} {\bibfnamefont {K.~Y.}\ \bibnamefont
  {Povarov}}, \bibinfo {author} {\bibfnamefont {S.~D.}\ \bibnamefont {Nabi}},
  \bibinfo {author} {\bibfnamefont {D.~G.}\ \bibnamefont {Mazzone}}, \bibinfo
  {author} {\bibfnamefont {J.}~\bibnamefont {Lass}}, \bibinfo {author}
  {\bibfnamefont {B.}~\bibnamefont {Roessli}}, \bibinfo {author} {\bibfnamefont
  {E.}~\bibnamefont {Ressouche}}, \bibinfo {author} {\bibfnamefont
  {Z.}~\bibnamefont {Yan}}, \bibinfo {author} {\bibfnamefont {S.}~\bibnamefont
  {Gvasaliya}},\ and\ \bibinfo {author} {\bibfnamefont {A.}~\bibnamefont
  {Zheludev}},\ }\href {https://doi.org/10.1103/PhysRevLett.129.087201}
  {\bibfield  {journal} {\bibinfo  {journal} {Phys. Rev. Lett.}\ }\textbf
  {\bibinfo {volume} {129}},\ \bibinfo {pages} {087201} (\bibinfo {year}
  {2022})}\BibitemShut {NoStop}%
\bibitem [{\citenamefont {Noginov}\ \emph {et~al.}(2001)\citenamefont
  {Noginov}, \citenamefont {Loutts}, \citenamefont {Ross}, \citenamefont
  {Grandy}, \citenamefont {Noginova}, \citenamefont {Lucas},\ and\
  \citenamefont {Mapp}}]{noginov_role_2001}%
  \BibitemOpen
  \bibfield  {author} {\bibinfo {author} {\bibfnamefont {M.~A.}\ \bibnamefont
  {Noginov}}, \bibinfo {author} {\bibfnamefont {G.~B.}\ \bibnamefont {Loutts}},
  \bibinfo {author} {\bibfnamefont {K.}~\bibnamefont {Ross}}, \bibinfo {author}
  {\bibfnamefont {T.}~\bibnamefont {Grandy}}, \bibinfo {author} {\bibfnamefont
  {N.}~\bibnamefont {Noginova}}, \bibinfo {author} {\bibfnamefont {B.~D.}\
  \bibnamefont {Lucas}},\ and\ \bibinfo {author} {\bibfnamefont
  {T.}~\bibnamefont {Mapp}},\ }\href {https://doi.org/10.1364/JOSAB.18.000931}
  {\bibfield  {journal} {\bibinfo  {journal} {Journal of the Optical Society of
  America B}\ }\textbf {\bibinfo {volume} {18}},\ \bibinfo {pages} {931}
  (\bibinfo {year} {2001})}\BibitemShut {NoStop}%
\bibitem [{\citenamefont {Sakakibara}\ \emph {et~al.}(1994)\citenamefont
  {Sakakibara}, \citenamefont {Mitamura}, \citenamefont {Tayama},\ and\
  \citenamefont {Amitsuka}}]{sakakibara_magnetometer_1994}%
  \BibitemOpen
  \bibfield  {author} {\bibinfo {author} {\bibfnamefont {T.}~\bibnamefont
  {Sakakibara}}, \bibinfo {author} {\bibfnamefont {H.}~\bibnamefont
  {Mitamura}}, \bibinfo {author} {\bibfnamefont {T.~T.~T.}\ \bibnamefont
  {Tayama}},\ and\ \bibinfo {author} {\bibfnamefont {H.~A.~H.}\ \bibnamefont
  {Amitsuka}},\ }\href {https://doi.org/10.1143/JJAP.33.5067} {\bibfield
  {journal} {\bibinfo  {journal} {Japanese Journal of Applied Physics}\
  }\textbf {\bibinfo {volume} {33}},\ \bibinfo {pages} {5067} (\bibinfo {year}
  {1994})}\BibitemShut {NoStop}%
\bibitem [{\citenamefont {K{\"u}chler}\ \emph {et~al.}(2023)\citenamefont
  {K{\"u}chler}, \citenamefont {Wawrzy{\'n}czak}, \citenamefont
  {Dawczak-D{\c{e}}bicki}, \citenamefont {Gooth},\ and\ \citenamefont
  {Galeski}}]{kuechler_new_applications_2023}%
  \BibitemOpen
  \bibfield  {author} {\bibinfo {author} {\bibfnamefont {R.}~\bibnamefont
  {K{\"u}chler}}, \bibinfo {author} {\bibfnamefont {R.}~\bibnamefont
  {Wawrzy{\'n}czak}}, \bibinfo {author} {\bibfnamefont {H.}~\bibnamefont
  {Dawczak-D{\c{e}}bicki}}, \bibinfo {author} {\bibfnamefont {J.}~\bibnamefont
  {Gooth}},\ and\ \bibinfo {author} {\bibfnamefont {S.}~\bibnamefont
  {Galeski}},\ }\href {https://doi.org/10.1063/5.0141974} {\bibfield  {journal}
  {\bibinfo  {journal} {Review of Scientific Instruments}\ }\textbf {\bibinfo
  {volume} {94}},\ \bibinfo {pages} {045108} (\bibinfo {year}
  {2023})}\BibitemShut {NoStop}%
\bibitem [{\citenamefont {K{\"u}chler}\ \emph {et~al.}(2017)\citenamefont
  {K{\"u}chler}, \citenamefont {W{\"o}rl}, \citenamefont {Gegenwart},
  \citenamefont {Berben}, \citenamefont {Bryant},\ and\ \citenamefont
  {Wiedmann}}]{kuechler_smallest_2017}%
  \BibitemOpen
  \bibfield  {author} {\bibinfo {author} {\bibfnamefont {R.}~\bibnamefont
  {K{\"u}chler}}, \bibinfo {author} {\bibfnamefont {A.}~\bibnamefont
  {W{\"o}rl}}, \bibinfo {author} {\bibfnamefont {P.}~\bibnamefont {Gegenwart}},
  \bibinfo {author} {\bibfnamefont {M.}~\bibnamefont {Berben}}, \bibinfo
  {author} {\bibfnamefont {B.}~\bibnamefont {Bryant}},\ and\ \bibinfo {author}
  {\bibfnamefont {S.}~\bibnamefont {Wiedmann}},\ }\href@noop {} {\bibfield
  {journal} {\bibinfo  {journal} {Review of Scientific Instruments}\ }\textbf
  {\bibinfo {volume} {8}},\ \bibinfo {pages} {083903} (\bibinfo {year}
  {2017})}\BibitemShut {NoStop}%
\bibitem [{\citenamefont {Starykh}\ \emph {et~al.}(2010)\citenamefont
  {Starykh}, \citenamefont {Katsura},\ and\ \citenamefont {Balents}}]{extreme}%
  \BibitemOpen
  \bibfield  {author} {\bibinfo {author} {\bibfnamefont {O.~A.}\ \bibnamefont
  {Starykh}}, \bibinfo {author} {\bibfnamefont {H.}~\bibnamefont {Katsura}},\
  and\ \bibinfo {author} {\bibfnamefont {L.}~\bibnamefont {Balents}},\ }\href
  {https://doi.org/10.1103/PhysRevB.82.014421} {\bibfield  {journal} {\bibinfo
  {journal} {Phys. Rev. B}\ }\textbf {\bibinfo {volume} {82}},\ \bibinfo
  {pages} {014421} (\bibinfo {year} {2010})}\BibitemShut {NoStop}%
\bibitem [{\citenamefont {Starykh}\ and\ \citenamefont
  {Balents}(2014)}]{nematic}%
  \BibitemOpen
  \bibfield  {author} {\bibinfo {author} {\bibfnamefont {O.~A.}\ \bibnamefont
  {Starykh}}\ and\ \bibinfo {author} {\bibfnamefont {L.}~\bibnamefont
  {Balents}},\ }\href {https://doi.org/10.1103/PhysRevB.89.104407} {\bibfield
  {journal} {\bibinfo  {journal} {Phys. Rev. B}\ }\textbf {\bibinfo {volume}
  {89}},\ \bibinfo {pages} {104407} (\bibinfo {year} {2014})}\BibitemShut
  {NoStop}%
\bibitem [{\citenamefont {Stoudenmire}\ and\ \citenamefont
  {Balents}(2008)}]{LB2008}%
  \BibitemOpen
  \bibfield  {author} {\bibinfo {author} {\bibfnamefont {E.~M.}\ \bibnamefont
  {Stoudenmire}}\ and\ \bibinfo {author} {\bibfnamefont {L.}~\bibnamefont
  {Balents}},\ }\href {https://doi.org/10.1103/PhysRevB.77.174414} {\bibfield
  {journal} {\bibinfo  {journal} {Phys. Rev. B}\ }\textbf {\bibinfo {volume}
  {77}},\ \bibinfo {pages} {174414} (\bibinfo {year} {2008})}\BibitemShut
  {NoStop}%
\end{thebibliography}
\end{document}